# Moyal Quantum Mechanics:

# The Semiclassical Heisenberg Dynamics

T. A. Osborn and F. H. Molzahn

Department of Physics
University of Manitoba
Winnipeg, MB, Canada, R3T 2N2

Abstract

The Moyal description of quantum mechanics, based on the Wigner–Weyl isomorphism between operators and symbols, provides a comprehensive phase space representation of dynamics. The Weyl symbol image of the Heisenberg picture evolution operator is regular in $\hbar$ and so presents a preferred foundation for semiclassical analysis. Its semiclassical expansion 'coefficients,' acting on symbols that represent observables, are simple, globally defined (phase space) differential operators constructed in terms of the classical flow. The first of two presented methods introduces a cluster-graph expansion for the symbol of an exponentiated operator, which extends Groenewold's formula for the Weyl product of two symbols and has $\hbar$ as its natural small parameter. This Poisson bracket based cluster expansion determines the Jacobi equations for the semiclassical expansion of 'quantum trajectories.' Their Green function solutions construct the regular $\hbar \downarrow 0$ asymptotic series for the Heisenberg–Weyl evolution map. The second method directly substitutes such a series into the Moyal equation of motion and determines the $\hbar$ coefficients recursively. In contrast to the WKB approximation for propagators, the Heisenberg–Weyl description of evolution involves no essential singularity in $\hbar$, no Hamilton–Jacobi equation to solve for the action, and no multiple trajectories, caustics or Maslov indices.

# I — Introduction

The Moyal phase space representation provides a self-contained and complete formulation of quantum mechanics, wherein the statement of dynamical evolution closely parallels that of classical Hamiltonian mechanics. In the Moyal formalism[1] each quantum observable $\hat{A}$ is represented, via the Wigner transform, by a unique real phase space function (symbol) $A_{\mathrm{w}}(q,p)$ and, conversely, each symbol defines an operator on Hilbert space via Weyl quantization. In this paper we analyze the Wigner–Weyl image of the Heisenberg picture evolution operator and through it obtain a semiclassical asymptotics of quantum evolution that is caustic free. These non-perturbative expansions are valid to arbitrary order in Planck's constant, are applicable to physically interesting observables, and hold for systems having a Hamiltonian symbol that is a smooth function on phase space.

Let $\hat{A}(t,s)$ be a quantum observable in the Heisenberg picture, for a system having an arbitrary time dependent Hamiltonian $\hat{H}(t)$. If $\sigma_{\mathrm{w}} : \hat{A} \mapsto A_{\mathrm{w}} = \sigma_{\mathrm{w}}(\hat{A})$ denotes the linear Wigner transform[2–8] (from operators in Hilbert space $\mathcal{H} = L^2(\mathbb{R}^d)$ to functions on the phase space $T^*\mathbb{R}^d$) then the evolution equation obeyed by the Weyl symbol $A_{\mathrm{w}}(t,\tau)$ is

$$\frac{\partial}{\partial \tau} A_{\mathrm{w}}(t,\tau) = -\{A_{\mathrm{w}}(t,\tau), H_{\mathrm{w}}(\tau)\}_{\mathrm{M}}. \tag{1.1}$$

Here the Moyal bracket $\{\ ,\ \}_{\mathrm{M}}$ is the image of the quantum commutator times $(i\hbar)^{-1}$ under the Wigner–Weyl isomorphism. When applied to (1.1) Bohr's correspondence principle is realized in a transparent fashion. The Moyal equation of motion (1.1) has the classical ($\hbar = 0$) limit

$$\frac{\partial}{\partial \tau} A(t,\tau) = -\{A(t,\tau), H_c(\tau)\}, \tag{1.2}$$

wherein $\{\ ,\ \}$ is the Poisson bracket. The fact that (1.2) is a transport equation for the classical dynamics governed by Hamiltonian function $H_c$ makes it easy to build $\hbar \downarrow 0$ expansions for quantum observables using the classical flow in phase space. Moreover, it turns out that equations (1.1), and (1.2), are naturally accompanied by initial conditions on $A_{\mathrm{w}}$, and $A$; this feature is responsible for the caustic-free nature of the attendant semiclassical expansion. We continue this introduction by providing a more detailed overview of this and other results in the paper.

In the usual Hilbert space operator based description of quantum mechanics, the unitary evolution $U(t,s)$ generated by $\hat{H}(t)$ is the operator valued solution of the Schrödinger initial value problem

$$i\hbar \frac{\partial}{\partial t} U(t,s) = \hat{H}(t) U(t,s), \qquad U(s,s) = \hat{I}; \tag{1.3}$$

$s$ is an arbitrary initial time, and $\hat{I}$ is the identity in $\mathcal{H}$. A Schrödinger picture observable $\hat{A}$ is transformed to the Heisenberg picture by the linear mapping $\boldsymbol{\Gamma}$ (from operators to operators) which conjugates $\hat{A}$ with $U$,

$$\boldsymbol{\Gamma}(s,t) : \hat{A} \mapsto U(t,s)^\dagger \hat{A} U(t,s) \equiv \hat{A}(t,s). \tag{1.4}$$

The Schrödinger operator $U(t,s)$ evolves states, whereas its Heisenberg counterpart $\boldsymbol{\Gamma}(s,t)$ evolves observables. Either of these two operators carries all dynamical information about the system and each is equally basic. However, the Heisenberg picture evolution operator $\boldsymbol{\Gamma}(s,t)$ has a much simpler semiclassical behavior than does $U(t,s)$. For this reason we focus our study largely on $\boldsymbol{\Gamma}(s,t)$. Good reasons for writing the time labels of $\boldsymbol{\Gamma}$ in the order $(s,t)$ will become apparent later.

Consider the ingredients that enter the Wigner–Weyl representation of $\boldsymbol{\Gamma}$.

As noted above, each observable $\hat{A}$ is identified with a function $A_w$ on phase space through the Wigner transform $\sigma_w$. (Subsection 2a reviews the Wigner–Weyl method.) An observable is called semiclassically admissible if its Weyl symbol is asymptotically regular at $\hbar = 0$,

$$A_w(\zeta) = A_c(\zeta) + \sum_{r=1}^\infty \frac{\hbar^r}{r!} a_r(\zeta). \tag{1.5}$$

Here the variable $\zeta = (q,p)$ denotes a generic point in the $2d$-dimensional phase space $\mathbb{R}^{2d} \simeq T^*\mathbb{R}^d$. The leading term of the series (1.5), $A_c(\zeta)$, is regarded as the classical counterpart of $\hat{A}$. It is easy to check that the common observables of quantum mechanics such as position, momentum, angular momentum, kinetic energy, the Hamiltonian and algebraic combinations of these operators are semiclassically admissible. If an observable depends on additional variables, such as the times $(t,s)$, then these labels are appended to all the functions in (1.5), e.g. $A_w$ becomes $A_w(t,s)$.

The classical analog of $U(t,s)$ is the measure preserving phase flow $g(t,s|\cdot) : \mathbb{R}^{2d} \to \mathbb{R}^{2d}$ which solves Hamilton's equation

$$\frac{d}{dt} g(t,s|\zeta) = J \nabla H_c\bigl(t, g(t,s|\zeta)\bigr) \tag{1.6}$$

with initial condition $g(s,s|\zeta) = \zeta \in \mathbb{R}^{2d}$. In (1.6) $J$ is the Poisson matrix carrying the symplectic structure of Hamilton's equation.

When the Heisenberg evolution $\boldsymbol{\Gamma}(s,t)$ is viewed in the Wigner–Weyl representation, it becomes the composition $\Gamma_{s,t} \equiv \sigma_w \circ \boldsymbol{\Gamma}(s,t) \circ \sigma_w^{-1}$. Stated less abstractly, an equivalent characterization of $\Gamma_{s,t}$ is the following: Suppose $A_w(t,\tau)$ is the solution of (1.1) with the $\tau = t$ initial value $A_w$; then

$$\Gamma_{s,t} : A_w \mapsto A_w(t,s) \tag{1.7a}$$

is the counterpart of (1.4) expressed as a linear mapping among phase space functions. The action of $\Gamma_{s,t}$ implements time evolution for the Weyl symbols of Heisenberg picture observables. We will write this mapping in the functional form

$$\Gamma_{s,t}(A_{\mathrm{w}}) = A_{\mathrm{w}}(t,s), \qquad (1.7b)$$

and refer to $\Gamma_{s,t}$ as the Heisenberg–Weyl evolution operator. A key feature of $\Gamma_{s,t}$ is that this mapping has a regular behavior with respect to Planck's constant $h = 2\pi\hbar$, and admits a small $\hbar$ formal asymptotic series

$$\Gamma_{s,t} = \sum_{n=0}^{\infty} \frac{\hbar^n}{n!} \gamma_{s,t}^{(n)} \qquad (1.8)$$

which is of a semiclassical nature. The leading term is the pull-back (on functions) of the classical phase flow, $\gamma_{s,t}^{(0)} = g(t,s|\,\cdot\,)^*$. Thus for an arbitrary time dependent function $a(t,\,\cdot\,)$ on phase space, $A(t,\tau,\,\cdot\,) \equiv \gamma_{\tau,t}^{(0)} a(t,\,\cdot\,) = a(t, g(t,\tau|\,\cdot\,))$ solves the transport equation (1.2).

Two different methods of computing the expansion (1.8) will be presented. The first method, found in §3, determines the evolved Weyl symbol of an arbitrary operator $\hat{A}(t)$ by first expressing it as a function of the canonical phase space coordinate operator $\hat{z} \equiv (\hat{q},\hat{p})$. The Heisenberg equation obeyed by $\hat{Z}(t,s) \equiv \boldsymbol{\Gamma}(s,t)\hat{z}$,

$$\frac{d}{dt}\hat{Z}(t,s) = J\nabla H_{\mathrm{w}}(t, \hat{Z}(t,s)), \qquad (1.9)$$

is a quantum version of Hamilton's equation (1.6). In (1.9) the necessary operator ordering prescription is understood to be Weyl ordering [see (3.6)]. The operator solution of (1.9) preserves the canonical commutation relations for all time; in Weyl symbol form:

$$\{Z_{\mathrm{w}}^i(t,s),\, Z_{\mathrm{w}}^j(t,s)\}_{\mathrm{M}} = J^{ij}. \qquad (1.10)$$

It is evident from (1.9) and (1.10) that it is appropriate to interpret $\hat{Z}(t,s)$ and its Weyl image $Z_{\mathrm{w}}(t,s)$ as the definition of a quantum trajectory. Furthermore, it will be shown that this quantum trajectory carries a generic semiclassical expansion whose leading term is the classical flow $g$.

With the aid of a new cluster-graph representation of the symbol of an exponentiated operator, it is possible to efficiently solve (1.9) for the the $\hbar$ expansion coefficients of $Z_{\mathrm{w}}(t,s)$. The higher order quantum fluctuations about the classical flow obey inhomogeneous Jacobi (variational) equations which may be solved recursively using a phase space Green function adapted to initial conditions. These basic ingredients are then used to construct the $\hbar$ expansion of $A_{\mathrm{w}}(t,s)$, from which it is straightforward to identify series (1.8).

Given the striking and simple similarities between the quantum and classical equations of motion (1.1) and (1.2), it is worth recalling some of the difficulties that have restricted the wider use of the Moyal formalism. The equation of motion (1.1) has an unpleasant non-local form. Specifically, when written in terms of phase space derivatives the bracket $\{\ ,\ \}_{\rm M}$ contains derivatives of arbitrarily high order. Thus (1.1) is a pseudodifferential equation in $2d+1$ variables. Similarly, the Heisenberg equation (1.9) for the canonical variables $\hat{Z}(t,s)$ is a $2d$ system of coupled ODE's which (for arbitrary Hamiltonians) is infinite-order non-linear with solutions that are non-commuting unbounded Hilbert space operators. These are substantial difficulties. A good portion of this paper is devoted to developing new calculational methods and representations that are effective in this environment.

The most novel of these representations is the above-mentioned cluster expansion, which provides a potent addition to the Wigner–Weyl calculus. Derived in §2b, the cluster expansion has the form

$$\left(e^{\hat{A}}\right)_{\rm w}(\zeta) = \exp\left\{\sum_{j=1}^{\infty} \frac{1}{j!} \mathcal{L}_j(\zeta;\hbar)\right\}, \qquad (1.11)$$

where the cluster functions $\mathcal{L}_j$ are determined by summation over simple connected graphs on $j$ vertices. Each vertex represents the symbol $A_{\rm w}$ and the graph edges are Poisson bracket operators coupling these symbols. Representation (1.11) has wide applicability. The case where $\hat{A}$ is a static (time independent) Hamiltonian and $e^{\hat{A}}$ is its evolution operator is discussed in §2d. The central role played by the exponential operator $e^{u\cdot\hat{z}}$ in the Wigner–Weyl method makes this a natural arena for the use of (1.11), as is demonstrated in §3.

A second method for computing $\Gamma_{s,t}$ is presented in §4. There, its Moyal equation

$$\frac{\partial}{\partial\tau}\Gamma_{\tau,t} = -\{\Gamma_{\tau,t}, H_{\rm w}(\tau)\}_{\rm M} \qquad (1.12)$$

is solved directly, order-by-order in $\hbar$, after (1.8) is substituted as an ansatz. The resulting recursive formulae for $\gamma^{(n)}_{s,t}$ are very compact and do not require previous computation of the $Z_{\rm w}(t,s)$ expansion. A comparison of the first few semiclassical expansion coefficients, generated by the two different approaches, shows that they are mutually consistent.

This second approach has antecedents[9–11] which consider the evolution of quantum densities (rather than observables) for static Hamiltonians. The difference between the Heisenberg equations for density and observable evolution is merely a sign. However, when time dependent observables and Hamiltonians are considered, as they are here, it turns out, somewhat unexpectedly, that the appropriate evolution problem results from the 'backward' rather than the 'forward' equation (in (1.1), $\partial/\partial\tau$ instead of $\partial/\partial t$), as is discussed in §4a.

There are two basic limitations imposed on the type of physical system considered here. First, it is assumed that the wave functions are scalar valued and consequently that the relevant quantum state space is $L^2(\mathbb{R}^d)$. Second, we presume that the classical phase space is flat $\mathbb{R}^{2d}$. Our results apply to systems characterized by smooth, semiclassically admissible Hamiltonian and observable symbols, both of which may be time dependent. Expansion (1.8) is constructible for $t$ in the maximal interval about $s$ for which the flow $g(t, s | \cdot)$ is defined on $\mathbb{R}^{2d}$. For smooth Hamiltonians with controlled $\zeta$-growth, the flow is globally defined for all time.

Although in principle one might obtain all our results by staying strictly within a framework intrinsic to the Weyl symbol calculus, we have found it useful to employ both the operator calculus and the symbol calculus, treating both points of view on an equal footing. The mathematical style of most of our arguments is heuristic and formal. It is designed to build, primarily by constructive means, the explicit semiclassical expansions found in Sections 3 and 4. A second generation approach to this subject would likely rewrite these results in the language of the Weyl pseudodifferential operators and Fourier integral operators. However in a first treatment these extra layers of mathematical rigor would tend to obscure the basic structure of the simple semiclassical asymptotics we find.

The Heisenberg–Weyl semiclassical expansion (1.8) has a number of distinctive and attractive features. The operator valued coefficients $\gamma_{s,t}^{(n)}$ are composed of the exact classical flow and phase space derivatives of finite order which act on the Weyl symbol of an arbitrary observable. The specific differential form of the coefficient is determined directly from the first $n+1$ cluster functions $\mathcal{L}_j$ without the need to solve any recurrence relations. If the Hamiltonian of the system is Weyl quantized the expansion has even $\hbar$-parity with all $\gamma_{s,t}^{(n)}$ for odd $n$ vanishing.

In §5 we discuss the computation of expectation values in the Heisenberg–Weyl formalism. The general features of this approach are illustrated by considering some simple examples. Appendix A reviews briefly the covariance of Weyl symbols under affine canonical transformations, and studies how this property is reflected in the exponential graph representations found in §3. Appendix B presents the connection between $\Gamma_{s,t}$ and its known integral kernel representations.

## II — Weyl Symbol Calculus

The formulation of the Weyl symbol calculus is reviewed in the first part of this section and stated in a manner suitable for the analysis of semiclassical dynamics. This summary also serves to fix our notations for quantum and classical phase space quantities. One cornerstone of this calculus is the $*$-product, which is the symbol image of the operator product. In §2b we develop a connected graph representation that determines the symbol for the exponential of operators, namely $\sigma_w\big(\exp(\hat{A})\big)$. This cluster-graph expansion is a natural extension of the Groenewold[12] asymptotic series for the $*$-product. Fourier transform techniques are employed in §2c to extend the cluster representation to describe $\sigma_w\big(f(\hat{A})\big)$ for suitably smooth functions $f$. As a sample application of these combinatorial graph identities, the Weyl symbol of the Schrödinger evolution operator $\exp(-it\hat{H}/\hbar)$ is discussed in §2d together with the appropriate realization of the Hamilton–Jacobi equation which serves as the basis of a WKB expansion for this propagator.

## 2a. The Wigner–Weyl Correspondence

In classical Hamiltonian mechanics, the state space of a system with configuration manifold $\mathbb{R}^d$ is the cotangent bundle $T^*\mathbb{R}^d$. The Cartesian chart $q$ on $\mathbb{R}^d$ induces canonical coordinates $\zeta = (q,p) = (q^1,\ldots,q^d,p_1,\ldots p_d)$ on this $2d$-dimensional phase space, which is then regarded as $\mathbb{R}^{2d} \simeq \mathbb{R}^d_q \times \mathbb{R}^d_p$. The geometry of phase space[13] is determined by the symplectic form $\omega = dp_i \wedge dq^i$. Of greater import here will be the Poisson structure determined by the matrix $J$ inverse to that of $\omega$. The Poisson bracket of two phase space functions $f$ and $g$ is the function $\{f,g\}$ defined by

$$\{f,g\}(\zeta) = \nabla f(\zeta) \cdot J \nabla g(\zeta), \qquad J = \begin{bmatrix} 0 & \delta \\ -\delta & 0 \end{bmatrix}, \tag{2.1}$$

where $\zeta$ now represents a generic point in $\mathbb{R}^{2d}$, and $\delta$ is the unit matrix (here $d$-dimensional).

Our discussion of the Weyl calculus in this section is not intended to be complete, regarding both the list of topics covered and the level of mathematical rigor. For example we do not discuss the statistical physics motivations which lead to the Wigner transform (cf. §4c and Refs. [1,2,7,12]), and we assume throughout that various formulae have meaning when the operators and symbols appearing in them are chosen from appropriate spaces.

For a variety of reasons, it is useful to associate to each operator $\hat{A}$ (in some class) a corresponding phase space function (perhaps generalized) called its symbol, and denoted $\sigma(\hat{A})$. There are many different ways[6,9,14,15] to set up such a symbol $\leftrightarrow$ operator correspondence. Most of these share a basic collection of common features. For example, it is reasonable to require a linear bijective correspondence, and that operator functions of $\hat{q}$ alone, or $\hat{p}$ alone, have as their symbols the same function of $q$, or $p$. The symbol of the $2d$-vector operator $\hat{z} = (\hat{q},\hat{p})$ will then be $\zeta$, understood as a linear polynomial. Acting on scalar wave functions $\psi \in L^2(\mathbb{R}^d)$, the $d$ position operators $\hat{q} = (\hat{q}^1,\ldots,\hat{q}^d)$ are defined by $\hat{q}^i \psi(x) = x^i \psi(x)$, and the conjugate momentum operators $\hat{p} = (\hat{p}_1,\ldots,\hat{p}_d)$ by $\hat{p}_j \psi(x) = -i\hbar \partial \psi / \partial x^j$.

Consider a less trivial function, $r(\zeta)$, that is a polynomial of degree two or greater. Then, generally, there are several different operator polynomials in $\hat{z}$ which reduce to $r(\zeta)$ if $\hat{z}$ is formally replaced by $\zeta$. They differ in their ordering of the non-commuting operators $\hat{q}^j$, $\hat{p}_j$. For a particular definition of the correspondence, only one of these operators can have $r$ as its symbol. This factor ordering aspect of the symbol ↔ operator correspondence is an unavoidable consequence of the canonical commutation relations

$$[\hat{z}^k, \hat{z}^l] = i\hbar J^{kl}. \tag{2.2}$$

The correspondence induced by normal ordering of $\hat{q}$ and $\hat{p}$ is familiar from classical PDE theory. There, if the operator $\hat{A}$ can be presented in the standard form

$$\hat{A} = \sum_{\beta \in \mathbb{W}^d} a_\beta(\hat{q}, \hbar) \hat{p}^\beta,$$

where $\beta$ is a $d$-multi-index and $\mathbb{W}$ the set of non-negative integers, then it is given the symbol $\sigma = \sigma_S$

$$\sigma_S(\hat{A})(\zeta) \equiv \sum_{\beta \in \mathbb{W}^d} a_\beta(q, \hbar) p^\beta,$$

which is called the 'standard' or '$q$-before-$p$'[9,16] symbol.

The Wigner–Weyl correspondence results from a more symmetrical choice of ordering of the components of $\hat{z}$, which we now summarize. A general phase space function $A(\zeta)$ is 'quantized' (roughly, $\zeta$ is replaced by $\hat{z}$) by defining '$A(\hat{z})$' according to the Weyl prescription:[17] in the Fourier resolution

$$A(\zeta) = \int_{\mathbb{R}^{2d}} du\, a(u)\, e^{iu \cdot \zeta} \tag{2.3a}$$

it is meaningful to replace $\zeta$ by $\hat{z}$, and the resulting operator

$$\hat{A} \equiv \int du\, a(u)\, e^{iu \cdot \hat{z}} \tag{2.3b}$$

may be regarded as $A(\hat{z})$. The Wigner–Weyl correspondence takes $A \leftrightarrow A(\hat{z})$ as an associated symbol ↔ operator pair. In particular, note that $e^{iu \cdot \zeta} \leftrightarrow e^{iu \cdot \hat{z}}$.

More precisely, the Weyl symbol of operator $\hat{A}$, denoted by $\sigma_w(\hat{A})$ or simply $A_w$, is defined as the Fourier transform, (2.3a), of the unique function (or distribution) $a$ for which representation (2.3b) holds. One should note at this stage that in general the symbol $A_w$ can possess a parametric dependence on $\hbar$.

Conversely, a given function $A_w$ will be the Weyl symbol of the operator

$$\hat{A} = (2\pi)^{-2d} \int du \int d\zeta\, A_w(\zeta)\, e^{iu \cdot (\hat{z} - \zeta \hat{I})}, \tag{2.4}$$

obtained by inverting (2.3a) and substituting into (2.3b). Formula (2.4) relates $\hat{A}$ to $A_w$ via integration. In formal calculations it is often very convenient to employ a differential form of this relation. This is gotten by writing $A_w(\zeta)e^{-iu\cdot\zeta} = [A_w(i\nabla_u)e^{-iu\cdot\zeta}]$ in (2.4) and performing the $\zeta$-integral. The result is

$$\hat{A} = A_w\left(\frac{1}{i}\nabla_u\right)e^{iu\cdot\hat{z}}\Big|_{u=0}. \tag{2.5}$$

Formula (2.5) clarifies the type of symmetrical ordering involved in Weyl quantization. Specifically, the basic symbol $A_w(\zeta) = \zeta^\alpha$ ($\alpha \in \mathbb{W}^{2d}$) is seen to be associated with the 'symmetric product' of operators[11] $(|\alpha|!)^{-1}\nabla_u^\alpha(u\cdot\hat{z})^{|\alpha|}$. For example, one has the association[18]

$$(q^j)^n(p_j)^m \leftrightarrow \frac{1}{2^n}\sum_{l=0}^{n}\binom{n}{l}(\hat{q}^j)^{n-l}(\hat{p}_j)^m(\hat{q}^j)^l.$$

Of course, an operator can be characterized by functions other than phase space-based symbols. A prime example is its integral kernel, i.e. the Dirac matrix element $\langle x|\hat{A}|y\rangle$, for which the following formulae are useful. Taking the $\langle x|\cdots|y\rangle$ matrix element of (2.4) leads to a construction of the kernel starting from the symbol,

$$\langle x|\hat{A}|y\rangle = h^{-d}\int dp\, e^{ip\cdot(x-y)/\hbar}A_w\left(\frac{x+y}{2},p\right). \tag{2.6}$$

(In deriving this one needs $\langle x|e^{i(u_1,u_2)\cdot\hat{z}}|y\rangle = e^{iu_1\cdot(x+y)/2}\delta(x-y+\hbar u_2)$.) One passes in the opposite direction, from kernel to symbol, via the Wigner transform[2]

$$A_w(q,p) = \int dv\, e^{-ip\cdot v/\hbar}\left\langle q+\frac{1}{2}v\,\Big|\,\hat{A}\,\Big|\,q-\frac{1}{2}v\right\rangle. \tag{2.7}$$

This familiar formula, often taken as the definition of the Weyl symbol, results from (2.6) by changing variables and inverting the Fourier transform. It follows readily from (2.7) that $\sigma_w(\hat{A}^\dagger) = \overline{\sigma_w(\hat{A})}$; in particular self-adjoint operators have real symbols.

If the symbol representation of quantum mechanics is to be advantageous, it is necessary to have a complete and practical symbol calculus. Of special relevance is the symbol of an operator product. For this basic structure there are again both integral- and derivative-based formulae which are useful in varying circumstances; for $z \in \mathbb{R}^{2d}$ they are

$$(\hat{A}\hat{B})_w(z) = (\pi\hbar)^{-2d}\int\int d\zeta d\zeta'\, A_w(z+\zeta)B_w(z+\zeta')\,e^{\frac{2i}{\hbar}\zeta\cdot J\zeta'}, \tag{2.8}$$

$$(\hat{A}\hat{B})_w(z) = \exp\left\{\frac{i\hbar}{2}\nabla_z\cdot J\nabla_{z'}\right\}A_w(z)B_w(z')\Big|_{z'=z}. \tag{2.9}$$

The two formulae (2.8) and (2.9) have a rather different mathematical status. The former constitutes a composition rule constructing the Weyl symbol of operator $\hat{A}\hat{B}$ in terms of the symbols of $\hat{A}$ and $\hat{B}$. Under rather general hypotheses[3,19−22] on the operators $\hat{A}$ and $\hat{B}$ this formula will be an exact, rigorous identity. In particular, (2.8) admits non-smooth symbols if the integration is understood in a distributional sense. The second statement — Groenewold's formula (2.9) — is a consequence of (2.8) and provides a small $\hbar$ expansion of $(\hat{A}\hat{B})_w$. Hörmander[20] and others[23–27] have attached various rigorous asymptotic interpretations to the expansion (2.9) by restricting both symbols $A_w$, $B_w$ to suitable spaces of smooth functions. For computational purposes, it is the algebraic and analytic form of the Groenewold formula that is its most important feature.

The right side of identity (2.8) may also be viewed as defining a bilinear, non-commutative, associative product of symbols. Denoted by $*$ and written

$$A_w * B_w \equiv (\hat{A}\hat{B})_w, \qquad (2.10)$$

this product is sometimes referred to as the star,[28,29] twisted[8,25,26] or Weyl[16] product. An elementary, but important, consequence of expansion (2.9) is that the $*$-product of two semiclassically admissible symbols is again semiclassically admissible.

Differences of operator products, i.e. commutators, are fundamental in quantum mechanics. Their image in the Wigner–Weyl formalism is the Moyal bracket $\{\,,\,\}_M$ defined by

$$i\hbar\{A_w, B_w\}_M \equiv A_w * B_w - B_w * A_w = [\hat{A}, \hat{B}]_w. \qquad (2.11)$$

Just like the commutator, the Moyal bracket is bilinear, skew and obeys the Jacobi identity. The Moyal bracket appearing in (1.1), for example, corresponds to the commutator in the Heisenberg equation of motion. Clearly, both the $*$-product and Moyal bracket operations on phase space functions are defined with reference to a specific value of the parameter $\hbar$.

To verify formulae (2.8) and (2.9) efficiently, begin by computing $\hat{A}\hat{B}$ using representations of the form (2.3$b$). After using the Baker–Campbell–Hausdorff (BCH) formula[30] to combine the exponential operators one has

$$\hat{A}\hat{B} = \int du \int dw\, a(u)b(w)\, e^{-\frac{i\hbar}{2} u \cdot Jw} e^{i(u+w)\cdot \hat{z}}, \qquad (2.12)$$

where $B_w$ is the Fourier transform of $b$, as in (2.3$a$). Note that the quantity $i\hbar J$ has arisen from the canonical commutator (2.2). Now, (2.12) expresses $\hat{A}\hat{B}$ in a form essentially parallel to (2.3$b$), so the relation analogous to (2.3$a$) is, immediately,

$$(\hat{A}\hat{B})_w(z) = \int du \int dw\, a(u)b(w)\, e^{-\frac{i\hbar}{2} u \cdot Jw} e^{i(u+w)\cdot z}. \qquad (2.13)$$

To obtain (2.8), substitute into (2.13) the Fourier representations of $a, b$ in terms of $A_w$, $B_w$ to find

$$(\hat{A}\hat{B})_w(z) = \int d\zeta \int d\zeta'\, A_w(\zeta) B_w(\zeta') \int\int \frac{du\, dw}{(2\pi)^{4d}}\, e^{iu\cdot(z-\zeta)} e^{iw\cdot(z-\zeta')} e^{-\frac{i\hbar}{2} u \cdot Jw}.$$

After shifting the $\zeta$ and $\zeta'$ integrations by $z$, one can do the $u$-integral to obtain $\delta(\zeta + \hbar Jw/2)$. This delta function trivializes the remaining $w$-integral and (2.8) results after minor rearrangement.

Next, to obtain (2.9) from (2.13), simply re-write the exponential factors in the integrand as
$$\exp\left\{\frac{i\hbar}{2}\nabla_z \cdot J\nabla_{z'}\right\} e^{i(u \cdot z + w \cdot z')}\bigg|_{z'=z}.$$
In the circumstances where the exponentiated derivative may be taken outside the $u$- and $w$-integrals of (2.13), these integrals reduce to $A_\mathrm{w}(z)$ and $B_\mathrm{w}(z')$, respectively [cf. (2.3a)]. This gives (2.9). It is interesting to note how the $\hbar$-coupled derivatives in (2.9) originate directly from quantum non-commutativity; if $\hat{q}$ and $\hat{p}$ were to commute then the symbol of $\hat{A}\hat{B}$ would be the ordinary product $A_\mathrm{w} B_\mathrm{w}$. It is the interchange of limiting order, related to pulling the differential operator $\nabla_z \cdot J\nabla_{z'}$ through the $du\,dw$ integration, that turns the result into an asymptotic expansion rather than an identity.

In many real applications it is not enough to consider the product of just two operators. Arbitrary products of operators, and functions of an operator, arise. Formula (2.9) is a basic first step for handling such cases. While the ensuing subsections will develop a more complete calculus, it is helpful to introduce here the concepts and notations which allow generalization of (2.9) to an arbitrary product.

Thus consider $n$ operators $A^{(l)}$ ($l \in \overline{n} \equiv \{1,\ldots,n\}$) with corresponding Weyl symbols $A_\mathrm{w}^{(l)}$. We introduce a product function defined on $n$ copies of phase space,

$$\underset{l\in\overline{n}}{\times} A_\mathrm{w}^{(l)} : (\mathbb{R}^{2d})^n \to \mathbb{C}, \qquad \underset{l\in\overline{n}}{\times} A_\mathrm{w}^{(l)}(\zeta_1,\ldots,\zeta_n) \equiv \prod_{l=1}^{n} A_\mathrm{w}^{(l)}(\zeta_l). \qquad (2.14)$$

In the case where all the symbols are the same, we often write simply $\times_{\overline{n}} A_\mathrm{w}$. If $n$ is small it is convenient to use a 'tuple' notation for the product function and write, for example, $\prec A_\mathrm{w}^{(1)}, A_\mathrm{w}^{(2)} \succ \equiv \times_{l\in\overline{2}} A_\mathrm{w}^{(l)}$.

Let $\nabla_j$ be the phase space gradient striking the $j^\mathrm{th}$ argument of this function,

$$\left[\nabla_j \underset{l\in\overline{n}}{\times} A_\mathrm{w}^{(l)}\right](\zeta_1,\ldots,\zeta_n) = [\nabla A_\mathrm{w}^{(j)}](\zeta_j) \prod_{l\neq j} A_\mathrm{w}^{(l)}(\zeta_l).$$

Notice that the $\nabla_j$'s mutually commute. In association with the symbol product $\times_{l\in\overline{n}} A_\mathrm{w}^{(l)}$ we introduce an extended Poisson bracket operator $B_{ij}$, which acts on the variables $\zeta_i$ and $\zeta_j$ and is defined by

$$B_{ij} = \nabla_i \cdot J\nabla_j. \qquad (2.15)$$

In this terminology the standard Poisson bracket (2.1) for a pair of functions $f$ and $g$ is recovered as $\{f,g\}(\zeta) = B_{12}\prec f, g\succ(\zeta)$ where the single argument $(\zeta)$ signifies the diagonal evaluation $(\zeta_1, \zeta_2)|_{\zeta_1=\zeta_2=\zeta}$. Restated with this new notation the Groenewold formula (2.9) reads

$$\left(A^{(1)} A^{(2)}\right)_{\mathrm{w}}(\zeta) = e^{\frac{i\hbar}{2} B_{12}} \prec A_{\mathrm{w}}^{(1)}, A_{\mathrm{w}}^{(2)} \succ (\zeta). \tag{2.16}$$

**Lemma 1.** The Weyl symbol of a product of $n \geq 2$ operators $\{A^{(l)}\}_{l=1}^{n}$ has the form

$$\left(A^{(1)} \ldots A^{(n)}\right)_{\mathrm{w}}(\zeta) = \exp\left\{\frac{i\hbar}{2} \sum_{1 \leq i < j \leq n} B_{ij}\right\} \underset{l \in \overline{n}}{\times} A_{\mathrm{w}}^{(l)}(\zeta). \tag{2.17}$$

**Proof:** If in (2.16) $A^{(2)}$ is replaced by $A^{(2)} A^{(3)}$ then

$$\left(A^{(1)} A^{(2)} A^{(3)}\right)_{\mathrm{w}}(\zeta) = e^{\frac{i\hbar}{2} \nabla_\zeta \cdot J \nabla_{\zeta'}} A_{\mathrm{w}}^{(1)}(\zeta) \left(A^{(2)} A^{(3)}\right)_{\mathrm{w}}(\zeta')\bigg|_{\zeta'=\zeta}$$

$$= e^{\frac{i\hbar}{2} \nabla_\zeta \cdot J \nabla_{\zeta'}} e^{\frac{i\hbar}{2} B_{23}} A_{\mathrm{w}}^{(1)}(\zeta) A_{\mathrm{w}}^{(2)}(\zeta') A_{\mathrm{w}}^{(3)}(\zeta')\bigg|_{\zeta'=\zeta}.$$

Now $\nabla_{\zeta'}$ reduces to $\nabla_2 + \nabla_3$ and (2.17) follows for the case $n = 3$, and for general $n$ by iterating this type of argument. ∎

The connection between the Moyal bracket, defined in (2.11), and the classical Poisson bracket is easily seen using (2.16). Since $J$ is skew, $B_{21} = -B_{12}$, and so for a pair of smooth phase space functions $A$ and $H$ one computes

$$i\hbar\{A, H\}_{\mathrm{M}} = e^{\frac{i\hbar}{2} B_{12}} \prec A, H \succ - e^{\frac{i\hbar}{2} B_{12}} \prec H, A \succ = \left(e^{\frac{i\hbar}{2} B_{12}} - e^{\frac{i\hbar}{2} B_{21}}\right) \prec A, H \succ,$$

$$\{A, H\}_{\mathrm{M}} = \frac{2}{\hbar} \sin\left(\frac{\hbar}{2} B_{12}\right) \prec A, H \succ = \sum_{l=0}^{\infty} \frac{(-)^l (\hbar/2)^{2l}}{(2l+1)!} B^{2l+1} \prec A, H \succ. \tag{2.18}$$

The semiclassical form of the Moyal bracket operator is thus $\{,\}_{\mathrm{M}} = \{,\} + \mathcal{O}(\hbar^2)$. This formal Poisson bracket limit is used in passing from (1.1) to (1.2). The symbol of the anti-commutator $\hat{A}\hat{H} + \hat{H}\hat{A}$ results from reversing the relative sign in (2.18), to give $2\cos\left(\frac{\hbar}{2} B_{12}\right) \prec A, H \succ$. The resulting $\cos + i \sin$ decomposition of the $*$-product forms[31] the basis of a superalgebra.

Another beneficial feature of the Wigner–Weyl calculus is its natural interaction with phase space translations, or more generally with affine canonical transformations. These properties are reviewed in Appendix A.

## 2b. Cluster Expansion of $\left(e^{\hat{A}}\right)_w$

Because of its central role in many calculations, it is necessary to have a general formula that determines the Weyl symbol of the exponential of an operator. We use a graph-combinatorial method to achieve this goal.

The following Lemma will be useful below. Notationally, $\sum_{\boldsymbol{m}}^{(n)}$ denotes summation over all multi-indices $\boldsymbol{m} = (m_1, \ldots, m_n) \in \mathbb{W}^n$ which partition the positive integer $n$, i.e. obey the constraint

$$\sum_{j=1}^{n} j m_j = n. \tag{2.19}$$

We also define $\sum_{\boldsymbol{m}}^{(0)} \equiv 1$.

**Lemma 2.** (Exponential Faà di Bruno) Let $\phi$ be a complex valued function of a real or complex variable, which is $C^n$ in the neighborhood of point $x$. Then

$$\left(\frac{d}{dx}\right)^n e^{\phi(x)} = e^{\phi(x)} \sum_{\boldsymbol{m}}^{(n)} c_{\boldsymbol{m}} \prod_{j=1}^{n} \left(\partial^j \phi(x)\right)^{m_j} \tag{2.20a}$$

where

$$c_{\boldsymbol{m}} \equiv n! \left[m_1! m_2! \cdots m_n! (1!)^{m_1} \cdots (n!)^{m_n}\right]^{-1}. \tag{2.20b}$$

**Proof:** This follows from the well-known Faà di Bruno formula for the $n^{\text{th}}$ derivative of a composite function. See Ref. [32]. ∎

The following graph terminology is required by the cluster expansion theorem below. The set $\mathcal{C}_j$ denotes all simple connected graphs[33] (clusters) which may be formed on the vertex set $\bar{\jmath} \equiv \{1, 2, \ldots, j\}$. Thus $C \in \mathcal{C}_j$ means $C = \{\bar{\jmath}, E\}$, where the edge set $E$ consists of unordered pairs $\alpha$ of distinct integers in $\bar{\jmath}$, linking a connected graph. A graph is simple if its edge set has at most one edge (or 'link') between each pair of vertex elements. A graph is connected if the edge set contains a unbroken path of links between each pair of vertex elements. The two integers comprising the edge $\alpha$ will be denoted $\wedge \alpha \equiv \min \alpha$ and $\vee \alpha \equiv \max \alpha$.

**Theorem 1.** Let $\hat{A}$ be an operator with Weyl symbol $A_w$. Then the symbol of $e^{\hat{A}}$ has the exponential cluster expansion

$$\left(e^{\hat{A}}\right)_w(\zeta) = \exp\left\{\sum_{j=1}^{\infty} \frac{1}{j!} \mathcal{L}_j(\zeta; \hbar)\right\} \tag{2.21a}$$

where $\mathcal{L}_j$ is the cluster function

$$\mathcal{L}_j(\zeta;\hbar) \equiv \sum_{C \in \mathcal{C}_j} \left[ \prod_{\alpha \in E} \left( e^{(i\hbar/2)B_\alpha} - 1 \right) \right] \underset{\bar{j}}{\times} A_{\mathrm{w}}(\zeta), \qquad (2.21b)$$

and the Poisson bracket operator for link $\alpha$ is defined by

$$B_\alpha \equiv \nabla_{\wedge\alpha} \cdot J \nabla_{\vee\alpha}. \qquad (2.21c)$$

**Proof:** Using formula (2.17) with each $A^{(l)} = \hat{A}$ yields

$$\left(e^{\hat{A}}\right)_{\mathrm{w}} = \sum_{n=0}^{\infty} \frac{1}{n!} (\hat{A}^n)_{\mathrm{w}} = \sum_{n=0}^{\infty} \frac{1}{n!} \prod_{1 \leq i < j \leq n} [1 + L_{ij}] \underset{\bar{n}}{\times} A_{\mathrm{w}} \qquad (2.22a)$$

where

$$L_{ij} \equiv e^{(i\hbar/2)B_{ij}} - 1. \qquad (2.22b)$$

Definition (2.22b) systematically splits the effect of the Groenewold operator $\exp[(i\hbar/2)B_{ij}]$ into its pure classical, $\hbar$ independent, piece (the identity) and the remaining non-classical contribution which is consolidated into $L_{ij}$. Since the operators $B_{ij}$ commute with one another so do the operators $L_{ij}$. The product $\prod_{i<j}[1 + L_{ij}]$ is a polynomial in $L_{ij}$ whose terms may be identified with simple graphs on $\bar{n} = \{1, \ldots, n\}$. Such a graph has edge $\{i, j\}$ if and only if link operator $L_{ij}$ appears in the term. In general this graph is a union of clusters (possibly trivial) on subsets of $\bar{n}$. If $I \subset \bar{n}$, define the cluster sum operator

$$S_I \equiv \sum_{C \in \mathcal{C}(I)} \prod_{\alpha \in E} L_\alpha, \qquad (2.23)$$

where $L_\alpha \equiv L_{\wedge\alpha,\vee\alpha}$. Let $\mathcal{P} = \{I_\lambda\}$ represent a partition of $\bar{n}$ into disjoint subsets $I_\lambda$. Then

$$\prod_{i<j}[1 + L_{ij}] = \sum_{\mathcal{P}} \prod_{I_\lambda \in \mathcal{P}} S_{I_\lambda}, \qquad (2.24)$$

the sum being taken over all possible partitions $\mathcal{P}$.

When (2.24) is applied to $\underset{\bar{n}}{\times} A_{\mathrm{w}}$, as required by (2.22a), then for each fixed $\mathcal{P}$ the result factors according to the subsets $I_\lambda$,

$$\prod_{i<j}[1 + L_{ij}] \underset{\bar{n}}{\times} A_{\mathrm{w}} = \sum_{\mathcal{P}} \prod_{I_\lambda \in \mathcal{P}} \left( S_{I_\lambda} \underset{I_\lambda}{\times} A_{\mathrm{w}} \right). \qquad (2.25)$$

The value of each factor here depends only upon the cardinality of $I_\lambda$. Thus consider the standard $j$-element ($1 \leq j \leq n$) subset $\bar{j}$ and define the $j^{\mathrm{th}}$ order cluster function in accord with (2.21b),

$$\mathcal{L}_j \equiv S_{\bar{j}} \underset{\bar{j}}{\times} A_{\mathrm{w}}. \qquad (2.26)$$

Then (2.25) can be written

$$\prod_{i<j}[1 + L_{ij}] \underset{\overline{n}}{\times} A_{\mathrm{w}} = \sum_{\mathcal{P}} \prod_{j=1}^{n} \mathcal{L}_j{}^{m_j(\mathcal{P})}, \qquad (2.27)$$

where $m_j(\mathcal{P})$ denotes the number of subsets $I_\lambda$ in $\mathcal{P}$ of cardinality $j$. Clearly $m_j = m_j(\mathcal{P})$ obeys (2.19).

Next, notice that the summand in (2.27) only depends on $\mathcal{P}$ via the 'occupation numbers' $\{m_j(\mathcal{P})\}$. Various distinct partitions possess the same multi-index $\boldsymbol{m}(\mathcal{P}) = \boldsymbol{m}$. For a given $\boldsymbol{m}$, the number of such partitions is found to be simply $c_{\boldsymbol{m}}$, defined in (2.20$b$). Combining this fact in (2.27) with (2.22$a$) gives

$$\left(e^{\hat{A}}\right)_{\mathrm{w}} = \sum_{n=0}^{\infty} \frac{1}{n!} \sum_{\boldsymbol{m}}{}^{(n)} c_{\boldsymbol{m}} \prod_{j=1}^{n} \mathcal{L}_j^{m_j}. \qquad (2.28)$$

The $n = 0$ term here is 1, as required by (2.22$a$).

Define the power series

$$\phi(x) = \sum_{j=1}^{\infty} \frac{x^j}{j!} \mathcal{L}_j$$

which reduces to the argument of the exponential in (2.21$a$) when $x = 1$. Clearly $\mathcal{L}_j = \partial^j \phi(0)$ and so application of the Faa di Bruno formula (2.20$a$) in (2.28) gives

$$\left(e^{\hat{A}}\right)_{\mathrm{w}} = \sum_{n=0}^{\infty} \frac{(1-0)^n}{n!} \left(\frac{d}{dx}\right)^n e^{\phi(x)} \bigg|_{x=0} = e^{\phi(1)},$$

by the Taylor series representation of $e^\phi$. ∎

The exponential cluster representation given in Theorem 1 is a simple and powerful computational tool, and is one of the main results of this paper. The combinatorial aspects of the proof of Theorem 1 are identical to those of the cluster expansions that are known for the statistical mechanical grand partition function,[34] and for the quantum propagators of perturbed quadratic Hamiltonians.[35] The algebraic framework of (2.21) is notably distinct from those applications of the cluster method, because it has no attendant integration for each symbol, and the basic binary coupling structure $B_{ij}$ is skew rather than symmetric.

Two properties of the cluster functions $\mathcal{L}_j$ should be pointed out now. First, it is clear that $\mathcal{L}_j$ is complex $j$-linear in the Weyl symbol $A_{\mathrm{w}}$. Second, notice that the link operator $L_\alpha = e^{(i\hbar/2)B_\alpha} - 1$ is formally $\mathcal{O}(\hbar^1)$, and that in (2.21$b$) each cluster $C$ (being connected) has at least $j - 1$ links $\alpha$. Thus

$$\mathcal{L}_j(\zeta; \hbar) = \mathcal{O}(\hbar^{j-1}) \qquad (2.29)$$

*provided* that $A_{\mathrm{w}} = \mathcal{O}(\hbar^0)$ (as is the case for semiclassically admissible observables $\hat{A}$).

The first three cluster functions are particularly simple,

$$\mathcal{L}_1(\zeta; \hbar) = A_{\mathrm{w}}(\zeta), \qquad \mathcal{L}_2(\zeta; \hbar) = L_{12} \prec A_{\mathrm{w}}, A_{\mathrm{w}} \succ, \qquad (2.30)$$

$$\mathcal{L}_3(\zeta; \hbar) = \{(L_{12}L_{23} + L_{13}L_{23} + L_{12}L_{13}) + (L_{12}L_{13}L_{23})\} \prec A_{\mathrm{w}}, A_{\mathrm{w}}, A_{\mathrm{w}} \succ (\zeta).$$

## 2c. The Symbol of $f(\hat{A})$

The results in §2b are tailored to exponential functions, but their scope can be broadened by introducing a generalized Fourier transform. Consider functions representable in the form

$$f(x) = \int d\mu(\alpha)\, e^{c\alpha x}, \qquad (2.31)$$

where $\mu$ is a real or complex measure on $\mathbb{R}$, and $c \in \mathbb{C}$. If (2.31) has meaning for $x$ in the spectrum of $\hat{A}$, one can define in a natural way the operator

$$f(\hat{A}) \equiv \int d\mu(\alpha)\, e^{c\alpha \hat{A}}. \qquad (2.32)$$

We now proceed formally and take the Wigner transform of (2.32) and employ Theorem 1,

$$f(\hat{A})_{\mathrm{w}}(\zeta) = \int d\mu(\alpha) \exp\left\{ \sum_{j=1}^{\infty} \frac{(c\alpha)^j}{j!} \mathcal{L}_j(\zeta;\hbar) \right\}. \qquad (2.33)$$

The explicit extraction of $(c\alpha)^j$ here means that $\mathcal{L}_j$ is the cluster function for $\hat{A}$ as defined by (2.21b). Since $\mathcal{L}_1 = A_{\mathrm{w}}$, (2.33) may be manipulated into

$$f(\hat{A})_{\mathrm{w}} = \int d\mu(\alpha) \exp\left\{ \sum_{j=2}^{\infty} \frac{\mathcal{L}_j}{j!} \left(\frac{\partial}{\partial y}\right)^j \right\} e^{c\alpha y}\bigg|_{y=A_{\mathrm{w}}} = \exp\left\{ \sum_{j=2}^{\infty} \frac{\mathcal{L}_j}{j!} \partial^j \right\} f(A_{\mathrm{w}}),$$

where $\partial$ is ordinary differentiation of $f$. A final useful rearrangement is to perform a cumulant expansion of the exponential, to obtain a series in higher derivatives of $f$. This is facilitated by Lemma 2 (with $x$ replaced by $\partial$), and the result is

$$f(\hat{A})_{\mathrm{w}}(\zeta) = f(A_{\mathrm{w}}(\zeta)) + \sum_{n=2}^{\infty} \sum_{\boldsymbol{m}\atop m_1=0}^{(n)} \left( \prod_{j=2}^{n} \frac{\mathcal{L}_j(\zeta;\hbar)^{m_j}}{(j!)^{m_j}\, m_j!} \right) \partial^n f(A_{\mathrm{w}}(\zeta)). \qquad (2.34)$$

This shows that the Weyl symbol of operator $f(\hat{A})$ is, to a first approximation, just the composition $f \circ A_{\mathrm{w}}$. The correction terms given by the $\sum_{n\geq 2}$ series turn out to be $\mathcal{O}(\hbar^2)$. To see this, recall that $\mathcal{L}_j = \mathcal{O}(\hbar^{j-1})$ and further notice that the order $\hbar^1$ contribution to $\mathcal{L}_2$ [cf. (2.30)] is the bracket $(i\hbar/2)\{A_{\mathrm{w}}, A_{\mathrm{w}}\}$, which vanishes.

## 2d. Schrödinger Evolution for Static Hamiltonians

As a first application of Theorem 1, it is natural to consider the time evolution of a quantum system governed by a static Hamiltonian $\hat{H}$. The Schrödinger evolution operator in this case is $U(t) = \exp(-it\hat{H}/\hbar)$. Its Weyl symbol is immediately expressed in exponential form,

$$U_{\rm w}(t) = \exp\Big\{\sum_{j=1}^{\infty} \mathcal{L}_j/j!\Big\}, \tag{2.35a}$$

where $\mathcal{L}_j$ is determined by (2.21b) with $A_{\rm w} = (t/i\hbar)H_{\rm w}$. From this it is seen that $\mathcal{L}_j \propto t^j$, and a cumulant expansion of (2.35a) will give the small $t$ Taylor series of $U_{\rm w}(t)$. Of course the exposed $\hbar^{-1}$ dependence of $A_{\rm w}$ means $U_{\rm w}(t)$ carries an essential singularity at $\hbar = 0$.

If one is interested in the WKB ($\hbar \downarrow 0$) asymptotic form (rather than $t \to 0$) then it is useful to expand the link operator thus,

$$L_\alpha = \sum_{l_\alpha=1}^{\infty} \frac{1}{l_\alpha!}\Big(\frac{i\hbar}{2}B_\alpha\Big)^{l_\alpha}.$$

There will be a sum over a link integer $l_\alpha$ for each edge $\alpha$, so introduce the extended graph summation notation

$$\sum_{\mathcal{G}_j} \equiv \sum_{C \in \mathcal{C}_j} \Big(\prod_{\alpha \in E}\sum_{l_\alpha=1}^{\infty}\Big)$$

and let $r \equiv \sum_{\alpha \in E} l_\alpha$ ($r = 0$ if $E = \emptyset$). Then the general cluster function formula is

$$\mathcal{L}_j(t,\zeta;\hbar) = t^j \sum_{\mathcal{G}_j} 2^{-r}(i\hbar)^{r-j} \prod_{\alpha \in E} \frac{1}{l_\alpha!} B_\alpha^{l_\alpha} \underset{\overline{j}}{\times} H_{\rm w}(\zeta). \tag{2.35b}$$

In particular for $j = 1, 2$:

$$\mathcal{L}_1 = \frac{t}{i\hbar} H_{\rm w}, \qquad \mathcal{L}_2 = \Big(\frac{t}{i\hbar}\Big)^2 \sum_{l=1}^{\infty} \frac{(i\hbar/2)^{2l}}{(2l)!} B_{12}^{2l} \prec H_{\rm w}, H_{\rm w} \succ.$$

In the second formula, all the odd terms vanished by skewness of the Poisson matrix.

It is also possible to investigate the Weyl symbol $U_{\rm w}(t)$ using WKB methods, in a fashion similar to the case of the familiar propagator $\langle x | U(t) | y \rangle$. The correct WKB ansatz turns out to be (at least for small $t$)

$$U_{\rm w}(t,\zeta) = \exp\big\{(i\hbar)^{-1}\Phi(t,\zeta;\hbar)\big\} \tag{2.36a}$$

$$\Phi(t,\zeta;\hbar) = \sum_{k=0}^{\infty} (i\hbar)^k \Phi_k(t,\zeta). \tag{2.36b}$$

Since $U(0) = \hat{I}$, thus $U_w(0) = 1$, and so one demands the initial conditions

$$\Phi_k(0, \zeta) = 0 \qquad (k \geq 0). \tag{2.37}$$

Briefly, the WKB method (which we won't carry out here) proceeds by substituting ansatz (2.36) into the Wigner transform of Schrödinger's equation, separating powers of $\hbar$ to find a recursive system of PDE's for $\{\Phi_k\}$, and solving them subject to (2.37) via a transport method.

The leading coefficient $\Phi_0(t, \zeta)$ controls the essential singularity in (2.36). It is the phase space analog[36−39] of the action $S(t, x, y)$ (Hamilton's principal function), and has been christened the 'phase action' by Marinov in his study.[40] The phase space Hamilton–Jacobi equation which $\Phi_0$ is required to obey via the WKB analysis is

$$\frac{\partial \Phi_0}{\partial t}(t, \zeta) = H_c\big(\zeta + \tfrac{1}{2} J \nabla \Phi_0(t, \zeta)\big). \tag{2.38}$$

Now, formulae (2.35) give a direct cluster-based construction of $U_w(t)$. Let us briefly compare its $\hbar$-structure with WKB ansatz (2.36). Consistency demands that $(i\hbar)^{-1}\Phi = \sum_{j \geq 1} \mathcal{L}_j/j!$. A cluster $C \in \mathcal{C}_j$ has at least $j - 1$ links and so the link integer sum $r \geq j - 1$. Thus, if $H_w = H_c + \mathcal{O}(\hbar^1)$, (2.35b) shows that each $\mathcal{L}_j = \mathcal{O}(\hbar^{-1})$, consistent with the structure of (2.36).

A next stage of comparison involves conjecturing a graphical formula for the phase action $\Phi_0$ by extracting the $(i\hbar)^{-1}$ parts of each $\mathcal{L}_j$. To do this, one must in (2.35b) replace $H_w$ by $H_c$ and keep only terms with $r = j - 1$. These correspond to minimally connected (tree) graphs, with each link integer $l_\alpha = 1$. The resulting series for the phase action is remarkably simple,

$$\Phi_0(t, \zeta) = t H_c(\zeta) + \sum_{j=3}^{\infty} \frac{t^j}{2^{j-1} j!} \sum_{C \in \mathcal{T}_j} \prod_{\alpha \in E} B_\alpha \underset{\overline{j}}{\times} H_c(\zeta). \tag{2.39}$$

Here $\mathcal{T}_j$ is the set of all tree graphs on $\overline{j}$, and the $j = 2$ term is again absent by skewness.

Note that the right side of (2.39) vanishes at $t = 0$ as required by (2.37). Although it is beyond our scope to present the details here, one can also verify that it satisfies the anticipated Hamilton–Jacobi equation, (2.38). Since that PDE is typically infinite-order nonlinear in $\Phi_0$, this is a nontrivial result.

Exponential cluster expansions of $U_w(t, s)$ for systems with non-static Hamiltonians of the form $p^2/2m + v(x, \tau)$ have been derived by Dyson series-based methods in Ref. [41]. Euclidean as well as toroidal configuration spaces were considered there.

# III — SEMICLASSICAL EVOLUTION OF OBSERVABLES

The first of two methods for computing semiclassical expansions for observable evolution in the Heisenberg picture and Weyl representation is presented here. We consider first the quantum trajectory $\hat{Z}(t,s)$. For this fundamental test case, the cluster expansion of Theorem 1 is used to derive Jacobi equations for the $\hbar$ expansion coefficients. These are solved in §3b using phase space Green function techniques. In the final subsection the expansion is extended to arbitrary observables.

Each specific quantum system is defined by a Hamiltonian operator $\hat{H}(t)$ which generates unitary evolution. Our derivations apply to any system described by a semiclassically admissible Hamiltonian; specifically, the Weyl symbol of $\hat{H}(t)$ is assumed to have the regular $\hbar \downarrow 0$ asymptotic expansion

$$H_{\mathrm{w}}(t, \hbar; \zeta) = H_c(t, \zeta) + \sum_{r=1}^{\infty} \frac{\hbar^r}{r!} h_r(t, \zeta). \tag{3.1}$$

Since $\hat{H}(t)$ is Hermitian $H_{\mathrm{w}}$, $H_c$ and $h_r$ are all real.

A simple but non-trivial example is that of a mass $m$ charged particle moving in an external electromagnetic field,

$$\hat{H}(t) = \frac{1}{2m} g^{\mu\nu} \left( \hat{p}_\mu - a_\mu(t, \hat{q}) \right) \left( \hat{p}_\nu - a_\nu(t, \hat{q}) \right) + \phi(t, \hat{q}).$$

Here $g^{\mu\nu}$ is a symmetric metric tensor and $(a, \phi)$ denote the vector and scalar potentials. Using (2.9) to compute the symbol of the quadratic portion of $\hat{H}(t)$ readily shows that $H_{\mathrm{w}}(t)$ is just the classical Hamiltonian $H_c(t, q, p)$ obtained by replacing $\hat{q}, \hat{p}$ with $q, p$ in the above expression. Thus no $h_r(t, \zeta)$ terms appear and $\hat{H}(t)$ is just the Weyl quantization of $H_c(t, \zeta)$. This particular Hamiltonian has a quadratic momentum dependence. In general, the Hamiltonian symbols in (3.1) allow for any smooth dependence in $p$.

The higher order $\hbar$ terms in (3.1) can originate via two inherently distinct mechanisms. The first one reflects the 'ordering ambiguity' in quantizing a classical observable. Different choices for ordering the non-commuting operators $\hat{q}^j, \hat{p}_j$ lead to different Hermitian operators consistent with a given classical $H_c(t, q, p)$, i.e. they all have $H_c$ as the $\hbar^0$ part of their Weyl symbol. But the higher $h_r$ coefficients differ because the commutation relations (2.2) permit one such operator to be expressed as another, plus discrepancy terms proportional to a factor of $i\hbar$ for each commutation. Thus, investigation of quantization schemes other than Weyl's will lead to Hamiltonians whose Weyl symbols contain nontrivial $h_r$.

The second mechanism is more intrinsic and concerns but a single operator: the true quantum Hamiltonian of the physical system of interest. There is no reason *a priori* why its Weyl symbol could not depend on $\hbar$. This need have nothing to do with ordering issues.

For example, $(1+\alpha\hbar)p^2/2m$ might be the free particle Hamiltonian, and it is ultimately up to experiment to detect such $\hbar$ dependence, if present.

Not all common observables are Weyl quantized. For example, the Hamiltonian for the quantum rigid rotator has the form $\hat{H} = \alpha\vec{L}^2$, for some constant $\alpha$ and $\vec{L} = \hat{q} \times \hat{p}$. The associated symbol is $H_{\mathrm{w}} = \alpha(q \times p)^2 - (3\alpha/2)\hbar^2$.

3a. *Semiclassical Quantum Trajectories*

The phase space coordinate operators $\hat{z} = (\hat{q}, \hat{p})$ form a basic set of observables. In the Heisenberg picture they become the time dependent operators $\hat{Z}(t,s) = \boldsymbol{\Gamma}(s,t)\hat{z}$, which as a consequence of (1.3) satisfy the Heisenberg equation of motion

$$i\hbar \frac{d}{dt}\hat{Z}(t,s) = \left[\hat{Z}(t,s),\, \hat{H}(t,s)\right] \tag{3.2a}$$

with initial condition

$$\hat{Z}(s,s) = \hat{z}, \tag{3.2b}$$

where the Hamiltonian in Heisenberg picture is $\hat{H}(t,s) = \boldsymbol{\Gamma}(s,t)\hat{H}(t)$.

For each fixed $s$, (3.2) constitutes an operator valued initial value problem which determines $\hat{Z}(t,s)$, if one supposes $\hat{H}(t,s)$ is given. We will seek a semiclassical description of $\hat{Z}(t,s)$ by making an ansatz for its Weyl symbol

$$Z_{\mathrm{w}}(t,s,\hbar;\zeta) = \sum_{r=0}^{\infty} \frac{\hbar^r}{r!}\, \mathbf{z}_r(t,s;\zeta). \tag{3.3}$$

Here the $\hbar$ dependence of the symbol $Z_{\mathrm{w}}(t,s)$ is explicitly displayed. The $\hbar$-free coefficient functions $\mathbf{z}_r(t,s;\zeta)$ in the $\hbar \downarrow 0$ formal asymptotic expansion (3.3) will be computed using techniques from classical mechanics. The initial conditions obeyed by these coefficients are a consequence of (3.2b). Specifically, $Z_{\mathrm{w}}(s,s,\hbar;\zeta) = \zeta$ and (3.3) imply

$$\mathbf{z}_0(s,s;\zeta) = \zeta, \qquad \mathbf{z}_r(s,s;\zeta) = 0 \quad (r > 0). \tag{3.4}$$

The Heisenberg equation in the form (3.2a) is, in practice, not well suited for solution. Note that its right side contains, in addition to $\hat{Z}(t,s)$, the Heisenberg operator $\hat{H}(t,s)$ which is unknown. One means of circumventing this difficulty, which we shall employ here, is to start from the following variant of (3.2a),

$$\frac{d}{dt}\hat{Z}(t,s) = \frac{1}{i\hbar}\boldsymbol{\Gamma}(s,t)\left[\hat{z},\, \hat{H}(t)\right]. \tag{3.5}$$

Here the Schrödinger commutator $\left[\hat{z},\, \hat{H}(t)\right]$ is given *a priori* and can be expressed as a known function of $\hat{z}$. Application of $\boldsymbol{\Gamma}(s,t)$ will then produce the same function of $\hat{Z}(t,s)$, thus rendering the Heisenberg equation in a suitable form. An alternative approach, used in §4, is to employ the 'backward' ($d/ds$) evolution equation instead of (3.2a).

The symbol of the commutator in (3.5) is readily computed because the symbol $z_{\mathrm{w}}(\zeta) = \zeta$ supports only one $\zeta$-derivative. Using (2.11) and (2.18)

$$\left[\hat{z}^j, \hat{H}(t)\right]_{\mathrm{w}} = i\hbar \left\{z_{\mathrm{w}}^j, H_{\mathrm{w}}(t)\right\}_{\mathrm{M}} = i\hbar\, B_{12} \prec z_{\mathrm{w}}^j, H_{\mathrm{w}}(t) \succ = i\hbar\, J^{jk} \nabla_k H_{\mathrm{w}}(t).$$

Now the commutator in (3.5) is represented in terms of this symbol via (2.5),

$$\frac{d}{dt}\hat{Z}(t,s) = J\nabla H_{\mathrm{w}}\!\left(t, \hbar; \frac{1}{i}\nabla_u\right) e^{iu \cdot \hat{Z}(t,s)}\bigg|_{u=0}. \tag{3.6}$$

This Heisenberg equation of motion was presented in the more stream-lined form (1.9) in order to emphasize its structural similarity to the classical Hamilton equation of motion.

To proceed with (3.6) we swap the roles of $iu$ and $(1/i)\nabla_u$ using the identity

$$f\!\left(\frac{1}{i}\nabla_u\right)g(u)\bigg|_{u=0} = g\!\left(\frac{1}{i}\nabla_u\right)f(u)\bigg|_{u=0}$$

which follows from Taylor expansion of the entire functions $f$ and $g$. This gives

$$\frac{d}{dt}\hat{Z}(t,s) = e^{\hat{Z}(t,s)\cdot\nabla} J\nabla H_{\mathrm{w}}(t,\hbar;0), \tag{3.7}$$

where all gradients $\nabla$ are understood to act on the (phase space) argument of $H_{\mathrm{w}}(t,\hbar;\cdot)$, with subsequent evaluation at $\zeta = 0$.

Finally, take the Wigner transform of the operator identity (3.7). (In doing so note that $\nabla$ is not a Hilbert space operator (such as $\hat{p}$), as it merely acts on the symbol $H_{\mathrm{w}}$ and can be treated like a fixed $\mathbb{C}^{2d}$-vector for purposes of the Wigner transform.) The cluster expansion, Theorem 1, is applicable if we choose the exponentiated operator to be $\hat{A} = \hat{Z}(t,s)\cdot\nabla$. Thus combining (3.7) with (2.21) leads to

$$\frac{d}{dt}Z_{\mathrm{w}}(t,s,\hbar;\zeta) = \exp\left\{\sum_{j=1}^{\infty} \frac{1}{j!}\mathcal{L}_j(\zeta;\hbar,t,s,\nabla)\right\} J\nabla H_{\mathrm{w}}(t,\hbar;0). \tag{3.8a}$$

Here the dependence of the cluster functions $\mathcal{L}_j$ upon $t$, $s$ and $\nabla$ is shown explicitly. Thus the $\mathcal{L}_j$ act as operators on the symbol $H_{\mathrm{w}}$ and are determined by

$$\mathcal{L}_j(\zeta;\hbar,t,s,\nabla) = \sum_{C\in\mathcal{C}_j} \prod_{\alpha\in E(C)} \left[e^{(i\hbar/2)B_\alpha} - 1\right] \underset{j}{\times} Z_{\mathrm{w}}(t,s,\hbar;\zeta)\cdot\nabla. \tag{3.8b}$$

The basic objective now is to substitute the $\hbar$-series ansatz (3.3) for $Z_{\mathrm{w}}$ into (3.8), and then to equate coefficients of common powers of $\hbar$ to obtain solvable relations for $\{\mathbf{z}_r\}_{r=0}^{\infty}$. In (3.8), $\hbar$-dependence not arising from $Z_{\mathrm{w}}$ resides in the link operators $L_\alpha = e^{(i\hbar/2)B_\alpha} - 1$, and possibly in the symbol $H_{\mathrm{w}}$. Note that the gradients in the operator $B_\alpha$ act only on the $Z_{\mathrm{w}}$ functions and not $H_{\mathrm{w}}$.

Consider the classical limit of (3.8) — i.e. set $\hbar = 0$. Then the link operators $L_\alpha$ vanish and so $\mathcal{L}_j = 0$, $j \geq 2$, because there is no contribution from any cluster having links. As for $\mathcal{L}_1$, equation (2.30) shows it to reduce to

$$\mathcal{L}_1(\zeta; 0, t, s, \nabla) = Z_{\mathrm{w}}(t, s, 0; \zeta) \cdot \nabla = \mathbf{z}_0(t, s; \zeta) \cdot \nabla.$$

The classical limit of (3.8) is therefore

$$\frac{d}{dt}\mathbf{z}_0(t, s; \zeta) = e^{\mathbf{z}_0(t,s;\zeta) \cdot \nabla} J\nabla H_c(t, 0) = J\nabla H_c\bigl(t, \mathbf{z}_0(t, s; \zeta)\bigr). \tag{3.9}$$

Equation (3.9) is precisely Hamilton's equation (1.6) for the classical mechanical system with Hamiltonian $H_c$. Its solution $\mathbf{z}_0$ as determined by the initial condition $\mathbf{z}_0(s, s; \cdot) = \mathrm{Id}_{\mathbb{R}^{2d}}$ [cf. (3.4)] is therefore the classical phase flow, $g(t, s|\zeta)$. That is,

$$\mathbf{z}_0(t, s; \cdot) = g(t, s|\cdot) : \mathbb{R}^{2d} \to \mathbb{R}^{2d}, \tag{3.10}$$

and, to leading order in $\hbar$ the quantum trajectory $Z_{\mathrm{w}}(t, s)$ is just the classical flow.

In order to derive the higher $\hbar$ corrections $\mathbf{z}_r$ from (3.8) it is useful to isolate this classical $g$-based underpinning, as follows. Let the deviation of $Z_{\mathrm{w}}$ from its classical limit $g$ be

$$\delta Z_{\mathrm{w}} \equiv Z_{\mathrm{w}} - g = \sum_{r=1}^{\infty} \frac{\hbar^r}{r!} \mathbf{z}_r(t, s; \zeta), \tag{3.11}$$

which is formally $\mathcal{O}(\hbar^1)$. Correspondingly, $\mathcal{L}_1$ splits into two parts,

$$\mathcal{L}_1(\zeta; \hbar, t, s, \nabla) = g(t, s|\zeta) \cdot \nabla + \delta Z_{\mathrm{w}}(t, s, \hbar; \zeta) \cdot \nabla,$$

the former of which will be used to shift the argument of $H_{\mathrm{w}}$ in (3.8), with the result

$$\frac{d}{dt}Z_{\mathrm{w}}(t, s, \hbar; \zeta) = e^{\delta Z_{\mathrm{w}}(t,s,\hbar;\zeta) \cdot \nabla} \exp\left\{\sum_{j=2}^{\infty} \frac{1}{j!}\mathcal{L}_j(\zeta; \hbar, t, s, \nabla)\right\} J\nabla H_{\mathrm{w}}\bigl(t, \hbar; g(t, s|\zeta)\bigr). \tag{3.12}$$

Equation (3.12) serves as a 'master equation' for extracting equations of motion for $\mathbf{z}_r$. This is done by collecting common powers of $\hbar$. The $\hbar$ structure on its left side is trivially obtained via (3.3). On the right side, each of the three factors is $\hbar$ dependent, with $e^{\delta Z_{\mathrm{w}} \cdot \nabla}$ and $J\nabla H_{\mathrm{w}}$ being relatively simple (cf. (3.11) and (3.1)).

Computing $\mathbf{z}_r$ requires analysis of $\mathcal{L}_j$ up to order $\hbar^r$. Recall from (2.29) that $\mathcal{L}_j(\zeta; \hbar) = \mathcal{O}(\hbar^{j-1})$; this remains true for $\mathcal{L}_j(\zeta; \hbar, t, s, \nabla)$ as given by (3.8b). The equations of motion for the first two quantum corrections, $\mathbf{z}_1$ and $\mathbf{z}_2$, will be extracted from (3.12). These examples also serve to expose the form of equation arising for general $r$. The $\hbar^1$ part of $\mathcal{L}_2$, and the $\hbar^2$ parts of $\mathcal{L}_2$ and $\mathcal{L}_3$ will therefore be needed for our calculation.

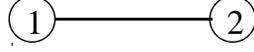

**Figure 1.** Cluster $C$ for $\mathcal{L}_2$

Consider first $\mathcal{L}_2$. In (3.8$b$) the only cluster in $\mathcal{C}_2$ is $C = \{\overline{2}, \{1,2\}\}$, which is illustrated in Fig. 1. So

$$\mathcal{L}_2 = \left[e^{(i\hbar/2)B_{12}} - 1\right] \prec Z_{\mathrm{w}} \cdot \nabla , \, Z_{\mathrm{w}} \cdot \nabla \succ$$
$$= \left(\frac{i\hbar}{2}\{Z_{\mathrm{w}}^j, Z_{\mathrm{w}}^k\} + \frac{1}{2}\left(\frac{i\hbar}{2}\right)^2 B_{12}{}^2 \prec Z_{\mathrm{w}}^j, Z_{\mathrm{w}}^k \succ\right)\nabla_j \nabla_k + \mathcal{O}(\hbar^3).$$

The first term vanishes because the $j,k$-skew matrix formed by Poisson bracket $\{Z_{\mathrm{w}}^j, Z_{\mathrm{w}}^k\}$ is contracted against the symmetric entity $\nabla_j \nabla_k$. Inserting expansion (3.3) for $Z_{\mathrm{w}}$ yields

$$\mathcal{L}_2 = -\frac{\hbar^2}{8} B_{12}^2 \prec g^j, g^k \succ \nabla_j \nabla_k + \mathcal{O}(\hbar^3). \tag{3.13}$$

In particular $\mathcal{L}_2$ has no $\hbar^1$ part, and so in (3.12) the sum $\sum_{j=2}^{\infty} \mathcal{L}_j/j! = \mathcal{O}(\hbar^2)$.

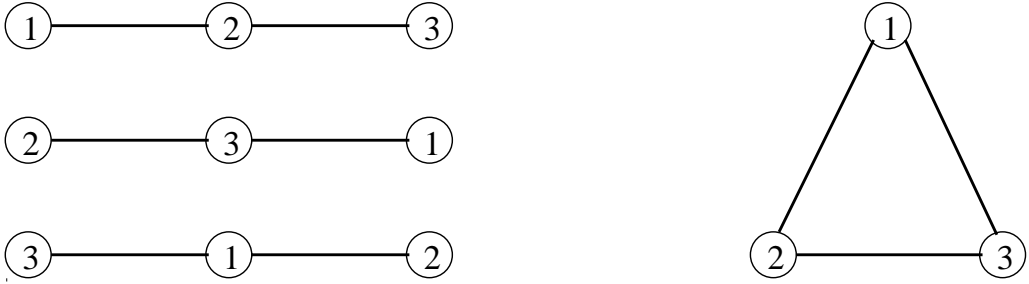

**Figure 2.** Clusters for $\mathcal{L}_3$

Turn now to the $\hbar^2$ part of cluster operator $\mathcal{L}_3$. The graphs contributing to $\mathcal{L}_3$, shown in Fig. 2, are three distinct labeled trees and the complete graph on $\overline{3}$. Since each $L_\alpha$ is $\mathcal{O}(\hbar^1)$, only the trees contribute to $\mathcal{L}_3$ at the lowest (second) order. Calculation of this term based on (3.8$b$) is straightforward; to within a sign each tree contributes equally and one finds that

$$\mathcal{L}_3 = \frac{\hbar^2}{4} B_{12} B_{23} \prec g^i, g^j, g^k \succ \nabla_i \nabla_j \nabla_k + \mathcal{O}(\hbar^3). \tag{3.14}$$

With these results we can now extract the $\hbar^1$- and $\hbar^2$-parts of (3.12). For $r = 1$,

$$\frac{d}{dt}\mathbf{z}_1(t,s;\zeta) = \bigl(\mathbf{z}_1(t,s;\zeta) \cdot \nabla\bigr) J\nabla H_c(t,g) + J\nabla h_1(t,g).$$

A natural and more suggestive rearrangement of this equation is

$$\mathcal{J}(t;s,\zeta)\mathbf{z}_1(t,s;\zeta) = J\nabla h_1\bigl(t, g(t,s|\zeta)\bigr), \tag{3.15}$$

where $\mathcal{J}$ is the Jacobi operator along $g$,

$$\mathcal{J}(t;s,\zeta) \equiv \frac{d}{dt} - J\nabla\nabla H_c\bigl(t, g(t,s|\zeta)\bigr). \tag{3.16}$$

($\nabla\nabla H_c$ denotes the Hessian matrix of $H_c$.) This Jacobi operator is associated with variational equations obtained by differentiating Hamilton's equation with respect to a parameter, such as $\zeta$. Thus (3.15) shows $\mathbf{z}_1$ to obey an inhomogeneous Jacobi equation (the inhomogeneous term $J\nabla h_1$ in this instance can actually be present only for non-Weyl quantized Hamiltonians).

Turning to the $r = 2$ case, one finds from (3.12)–(3.14) that $\mathbf{z}_2$ also obeys such a Jacobi equation,

$$\mathcal{J}(t;s,\zeta)\mathbf{z}_2 = \left[(\mathbf{z}_1 \cdot \nabla)^2 - \frac{1}{8}B_{12}^2 \underset{2}{\times} g \cdot \nabla + \frac{1}{12}B_{12}B_{23} \underset{3}{\times} g \cdot \nabla\right] J\nabla H_c(t,g)$$
$$+ 2(\mathbf{z}_1 \cdot \nabla) J\nabla h_1(t,g) + J\nabla h_2(t,g). \tag{3.17}$$

In this instance, the inhomogeneous term need not vanish in the Weyl quantized case.

The general features of equations (3.15) and (3.17) turn out to be generic for $r \geq 1$. To summarize these features, $\mathbf{z}_r$ obeys an inhomogeneous Jacobi equation

$$\mathcal{J}(t;s,\zeta)\mathbf{z}_r = f_r(t,s;\zeta). \tag{3.18}$$

The Jacobi operator $\mathcal{J}$ in (3.18) arises by considering the only two terms in the $\hbar^r$ part of (3.12) that contain $\mathbf{z}_r$: $d/dt$ comes from the left side and $J\nabla\nabla H_c$ originates in the linear $\delta Z_{\mathrm{w}} \cdot \nabla$ term in the expansion of $e^{\delta Z_{\mathrm{w}} \cdot \nabla}$. The inhomogeneous term $f_r$ is built from $g, \mathbf{z}_1, \ldots, \mathbf{z}_{r-1}$ and from $H_c, h_1, \ldots, h_r$. The next subsection will prove solvability of (3.18) when $f_r$ is known. Since $h_i$ are given from the quantum Hamiltonian, this implies the possibility of solving the $r$-indexed family (3.18) recursively for $\{\mathbf{z}_r\}$.

## 3b. Jacobi Green Function Solutions

In this subsection we explicitly construct the solution of the inhomogeneous Jacobi equation (3.18) subject to the relevant homogeneous initial condition (3.4), i.e.

$$\mathbf{z}_r(s,s;\zeta) = 0, \qquad r \geq 1. \tag{3.19}$$

Equation (3.18) is a $2d$-dimensional first order ODE, with parameter $\zeta \in \mathbb{R}^{2d}$. Basic ODE theory implies that the initial value problem (3.18)–(3.19) has a unique solution $\mathbf{z}_r$, provided the given coefficient functions $f_r$ and $\nabla\nabla H_c(t,g)$ are sufficiently smooth in $t$.

This problem can be solved by constructing the phase space Green function tailored to it. The *Green function* $G$ for (3.18)–(3.19) is a matrix function $G : \mathbb{R} \times \mathbb{R} \to \mathbb{R}^{2d \times 2d}$ obeying the fundamental inhomogeneous Jacobi equation

$$\mathcal{J}(t;s,\zeta)G(t,\rho) = \delta(t-\rho)\,J \qquad (\rho \in \mathbb{R}), \tag{3.20}$$

and the homogeneous initial condition

$$G(s,\rho) = 0 \qquad (\rho \in \mathbb{R}). \tag{3.21}$$

The necessary dependence of $G$ upon the parameters $(s,\zeta)$ appearing in $\mathcal{J}$ is being notationally suppressed. Once the Green function is known one verifies directly that the solution to (3.18)–(3.19) is

$$\mathbf{z}_r(t,s;\zeta) = \int_{\mathbb{R}} d\rho\, G(t,\rho) J^{-1} f_r(\rho,s;\zeta). \tag{3.22}$$

The construction of $G$ based on phase space Jacobi fields is relatively straightforward. Recall that a *Jacobi field*, $B$, is a matrix solution of the homogeneous Jacobi equation; namely, $B : \mathbb{R} \to \mathbb{R}^{2d \times m}$ satisfies

$$\mathcal{J}(t;s,\zeta)B(t) = 0.$$

Thus each of $B$'s $m$ columns is a vector Jacobi field.

**Lemma 3.** Let $B$ be any $2d \times 2d$ matrix Jacobi field which is invertible at some time $\rho_o$. Then $B(\rho)$ is invertible for all $\rho \in \mathbb{R}$ and

$$G(t,\rho) \equiv \bigl[\Theta(t-\rho) - \Theta(s-\rho)\bigr] B(t) B(\rho)^{-1} J \tag{3.23}$$

is an initial value Green function obeying (3.20)–(3.21).

**Proof:** A basic property of Jacobi fields is symplectic conservation: if $A$ and $B$ are, respectively, $2d \times m$ and $2d \times m'$ Jacobi fields, with $A^{\mathrm{T}}$ the transpose of $A$, then

$$A(t)^{\mathrm{T}} J B(t) = \mathrm{const.} \in \mathbb{R}^{m \times m'}.$$

This is readily verified by $t$-differentiation. Choose $A = B$ here and take the determinant to find $\left(\det B(t)\right)^2$ is a constant, which is nonzero by considering $t = \rho_o$. Hence $B$ remains invertible. Next, it is immediate that (3.23) obeys (3.21), and (3.20) follows by a simple calculation. ∎

To complete our construction of $G$ a suitable Jacobi field is needed. A natural choice is to use the derivative of the phase flow

$$B(t) = B(t; s, \zeta) \equiv \nabla g(t, s | \zeta) = \frac{\partial g}{\partial \zeta}(t, s | \zeta). \tag{3.24}$$

(The standard derivative–matrix notation is used here: $\partial g / \partial \zeta$ is the $2d$ square matrix with $(i, j)$ entry $\partial g^i / \partial \zeta^j$.) That $\nabla g$ is a Jacobi field for $\mathcal{J}(t; s, \zeta)$ follows by taking the $\zeta$-derivative of Hamilton's equation for $g$. Moreover at time $\rho_o = s$, $\nabla g(s, s | \zeta) = \delta$ is invertible. We also note that this choice of $B(t; s, \zeta)$ is symplectic (i.e. $B^{\mathrm{T}} J B = J$), hence

$$B(\rho)^{-1} = J^{-1} B(\rho)^{\mathrm{T}} J. \tag{3.25}$$

To summarize, the solution of (3.18)–(3.19) gotten by combining (3.22)–(3.25) is

$$\mathbf{z}_r(t, s; \zeta) = \int_s^t d\rho \, \nabla g(t, s | \zeta) J \nabla g(\rho, s | \zeta)^{\mathrm{T}} J^{-1} f_r(\rho, s; \zeta). \tag{3.26}$$

It should be emphasized that there are no restrictions on allowed values of $t, s$ here, apart from the requirement that the classical flow be defined. This is a result of the fact that *initial value* problems are being solved. This in turn is not just a consequence of the initial value nature of Schrödinger's equation, but also reflects the fact that phase space symbols of Heisenberg picture operators are being studied. If one alternatively studies, for example, the conventional propagator $\langle x | U(t, s) | y \rangle$ or the Weyl symbol $U_{\mathrm{w}}(t, s)$, then as is well known the relevant trajectories and Green functions obey boundary conditions involving separate times $t$ and $s$ (in place of initial condition (3.21)). Such two-time boundary problems do not always have solutions; specifically, this occurs at conjugate points. The description of evolution being considered here is, by contrast, free of such caustics.

We now apply the general formula (3.26) to solve in closed form the Jacobi equations for $\mathbf{z}_1$ and $\mathbf{z}_2$ derived in the previous subsection. The right side of (3.15) defines $f_1(t, s; \zeta)$ and so

$$\mathbf{z}_1(t, s; \zeta) = \int_s^t d\rho \, \nabla g(t, s | \zeta) J \nabla g(\rho, s | \zeta)^{\mathrm{T}} \nabla h_1(\rho, g(\rho, s | \zeta)). \tag{3.27}$$

This simplifies by using the chain rule to express $\nabla g^{\mathrm{T}} \nabla h_1(\rho, g) = \nabla_\zeta h_1(\rho, g(\rho, s|\zeta))$, and recognizing the resulting integrand of (3.27) to be a Poisson bracket. Thus

$$\mathbf{z}_1(t,s;\zeta) = \int_s^t d\rho \, \{ g(t,s|\cdot) \,, \, h_1(\rho, g(\rho, s|\cdot)) \}(\zeta). \tag{3.28}$$

For $\mathbf{z}_2$ we shall restrict attention to the usual case of a Weyl quantized Hamiltonian, $H_{\mathrm{w}} = H_c$. In this case $h_1$, $h_2$ and $\mathbf{z}_1$ are all zero, and combining (3.17) with (3.26) gives

$$\mathbf{z}_2(t,s;\zeta) = \int_s^t d\rho \, \nabla g(t,s|\zeta) J \nabla g(\rho, s|\zeta)^{\mathrm{T}} \Big[ -\frac{1}{8} B_{12}{}^2 \prec g^\alpha(\rho, s)\,, g^\beta(\rho, s) \succ (\zeta) \nabla_\alpha \nabla_\beta$$
$$+ \frac{1}{12} B_{12} B_{23} \prec g^\alpha(\rho, s)\,, g^\beta(\rho, s)\,, g^\gamma(\rho, s) \succ (\zeta) \nabla_\alpha \nabla_\beta \nabla_\gamma \Big] \nabla H_c(\rho, g(\rho, s|\zeta)). \tag{3.29}$$

### 3c. Evolution of Semiclassically Admissible Observables

The prior results of this section have succeeded in developing a recursive scheme for computing a semiclassical expansion of the quantum trajectory $\hat{Z}(t,s)$. Here we extend those results to the case of more arbitrary observables $\hat{A}(t,s)$ in a simple way.

Let $\hat{A}(t)$ be a time dependent Schrödinger picture operator which is semiclassically admissible. Then its symbol has the form, fully parallel to (3.1),

$$A_{\mathrm{w}}(t, \hbar; \zeta) = A_c(t, \zeta) + \sum_{r=1}^\infty \frac{\hbar^r}{r!} a_r(t, \zeta). \tag{3.30}$$

The $\hbar$-expansion coefficients $A_c$ and $a_r$ are real if $\hat{A}(t)$ is self-adjoint. As the notation in (3.30) suggests, $A_c(t, \zeta)$ is the classical, $\hbar$ independent, part of $A_{\mathrm{w}}(t, \hbar; \zeta)$.

In the Heisenberg picture, the evolution of $\hat{A}(t)$ is $\hat{A}(t,s) = \boldsymbol{\Gamma}(s,t) \hat{A}(t)$. We will construct the Weyl symbol of $\hat{A}(t,s)$ as an expansion of the form

$$A_{\mathrm{w}}(t,s,\hbar;\zeta) = \sum_{r=0}^\infty \frac{\hbar^r}{r!} \mathbf{a}_r(t,s;\zeta) \tag{3.31}$$

under the assumption that $\mathbf{z}_k(t,s;\zeta)$ are known. The double time labels in $A_{\mathrm{w}}(t,s,\hbar;\zeta)$ and $\mathbf{a}_r(t,s;\zeta)$ distinguish these symbols as dynamical objects, in contrast to the non-dynamical quantities $A_{\mathrm{w}}(t,\hbar;\zeta)$ and $a_r(t,\zeta)$ bearing single time arguments in (3.30). In particular, (3.31) shows that semiclassical admissibility is preserved under evolution.

Using methods already employed in §3a we directly compute

$$\hat{A}(t,s) = A_{\mathrm{w}}\left(t,\hbar; \frac{1}{i}\nabla_u\right) \boldsymbol{\Gamma}(s,t) e^{iu \cdot \hat{z}} \Big|_{u=0} = A_{\mathrm{w}}\left(t,\hbar; \frac{1}{i}\nabla_u\right) \exp\{iu \cdot \hat{Z}(t,s)\} \Big|_{u=0}$$
$$= \exp\{\hat{Z}(t,s) \cdot \nabla\} A_{\mathrm{w}}(t,\hbar; 0). \tag{3.32}$$

Taking the Wigner transform, using the cluster expansion theorem and isolating the classical flow yields

$$A_{\text{w}}(t,s,\hbar;\zeta) = e^{\delta Z_{\text{w}}(t,s,\hbar;\zeta) \cdot \nabla} \exp\Big\{ \sum_{j=2}^{\infty} \frac{1}{j!} \mathcal{L}_j(\zeta;\hbar,t,s,\nabla) \Big\} A_{\text{w}}\big(t,\hbar; g(t,s|\zeta)\big). \qquad (3.33)$$

In the above formulae $\nabla$ is understood to act on the symbol $A_{\text{w}}(t,\hbar;\,\cdot\,)$. The cluster operator $\mathcal{L}_j$ contains $j$ such gradients. Hence if $A_{\text{w}}(t,\hbar;\zeta)$ is polynomial in $\zeta$ the $\sum_{j=2}^{\infty} \mathcal{L}_j/j!$ series will reduce to a finite sum, and the expansion of $e^{\delta Z_{\text{w}} \cdot \nabla}$ will similarly terminate. This kind of simplification underscores the fact that most of the nontrivial dynamical effects are built into $Z_{\text{w}}(t,s,\hbar;\zeta)$.

Formula (3.33) verifies the $\hbar$-analytic structure of the evolving symbol $A_{\text{w}}(t,s)$ which was asserted in (3.31). Moreover, by extracting the coefficients of common powers of $\hbar$ in (3.33) one obtains explicit formulae for $\mathbf{a}_r(t,s;\zeta)$ in terms of the known quantities $\{\mathbf{z}_k(t,s;\zeta), a_k(t,\zeta)\}_{k=0}^{r}$. The first three of these are found to be

$$\mathbf{a}_0(t,s;\zeta) = A_c\big(t, g(t,s|\zeta)\big), \qquad (3.34)$$

$$\mathbf{a}_1(t,s;\zeta) = a_1\big(t, g(t,s|\zeta)\big) + \mathbf{z}_1(t,s;\zeta) \cdot \nabla A_c\big(t, g(t,s|\zeta)\big), \qquad (3.35)$$

$$\mathbf{a}_2(t,s;\zeta) = a_2\big(t, g(t,s|\zeta)\big) + 2\mathbf{z}_1(t,s;\zeta) \cdot \nabla a_1\big(t, g(t,s|\zeta)\big) + \Big[\big(\mathbf{z}_1(t,s;\zeta) \cdot \nabla\big)^2 \qquad (3.36)$$
$$+ \mathbf{z}_2(t,s;\zeta) \cdot \nabla - \frac{1}{8} B_{12}{}^2 \underset{2}{\times} g(t,s|\zeta) \cdot \nabla + \frac{1}{12} B_{12} B_{23} \underset{3}{\times} g(t,s|\zeta) \cdot \nabla\Big] A_c\big(t, g(t,s|\zeta)\big).$$

Clearly this method of determining $\mathbf{a}_r$ requires no further integration of ODE's beyond that needed to determine the $\mathbf{z}_r$ coefficients. In this sense $\{\mathbf{z}_r\}$ act as a 'basis' with which to represent arbitrary evolving observables.

Representations of the $\hbar$-coefficients $\gamma_{s,t}^{(n)}$ of the Heisenberg–Weyl evolution map $\Gamma_{s,t}$ are obtained from (3.33) by choosing the observable symbol to be a static $\hbar$ independent phase space function, say $a$. Then $a_r(t,\,\cdot\,) = \delta_{r,0}\, a$ and

$$\gamma_{s,t}^{(0)}(a) = a\big(g(t,s|\,\cdot\,)\big) = g(t,s)^*\, a, \qquad (3.37)$$

while from (3.35)
$$\gamma_{s,t}^{(1)}(a) = \mathbf{z}_1(t,s;\,\cdot\,) \cdot \nabla a\big(g(t,s|\,\cdot\,)\big). \qquad (3.38)$$

For a Weyl quantized system, (3.36) leads to

$$\gamma_{s,t}^{(2)}(a) = \mathbf{z}_2(t,s;\,\cdot\,) \cdot \nabla a\big(g(t,s|\,\cdot\,)\big) - \frac{1}{8} B_{12}{}^2 \prec g^\alpha(t,s),\, g^\beta(t,s) \succ \nabla_\alpha \nabla_\beta a\big(g(t,s|\,\cdot\,)\big)$$
$$+ \frac{1}{12} B_{12} B_{23} \prec g^\alpha(t,s),\, g^\beta(t,s),\, g^\gamma(t,s) \succ \nabla_\alpha \nabla_\beta \nabla_\gamma a\big(g(t,s|\,\cdot\,)\big). \qquad (3.39)$$

Note that, by the linearity of $\Gamma_{s,t}$, in order to compute $A_{\mathrm{w}}(t,s)$ it is really sufficient to determine just $\{\gamma_{s,t}^{(n)} a\}$. For then (3.31) follows from (3.30) and (1.8) with

$$\mathbf{a}_r(t,s;\,\cdot\,) = \sum_{n=0}^{r} \binom{r}{n} \gamma_{s,t}^{(n)}\, a_{r-n}(t,\,\cdot\,).$$

This approach will be taken in the next section.

The derivation leading to (3.33) is formally exact — no approximations have been made nor terms neglected. The result provides a representation of evolution which explicitly exposes the underlying classical motion and systematically characterizes higher order quantum corrections in terms of a cluster-graph expansion. Representation (3.33) is universal in the sense that it holds for all Hamiltonians $\hat{H}(t)$ and observables $\hat{A}(t)$ that are semiclassically admissible operators. Identity (3.34) states that to leading order in $\hbar$ the evolving Weyl symbol $A_{\mathrm{w}}(t,s,\hbar;\zeta)$ is given by the classical observable pulled along the classical flow. This is the analog, for Weyl symbols, of Egorov's Theorem.[42] In (3.33) the dynamical quantum corrections to classical behavior are consolidated into two aspects: 1) $\delta Z_{\mathrm{w}}$ which accounts for the deviation of the quantum trajectory $Z_{\mathrm{w}}$ about its classical limit $g$, and 2) the $\hbar$ dependence of the cluster operators $\mathcal{L}_j$ which has its origin in the $*$-product.

## IV — MOYAL BRACKET METHOD

Here a pure classical transport recurrence relation method of computing the semiclassical expansion (3.31) is constructed. This alternative approach does not rely on the $Z_{\mathrm{w}}$ representation (3.3) or the cluster expansion of §2b. The transport based methods of this section are an extension of techniques found in the prior literature[9−11,39,43] on semiclassical expansions rooted in the Moyal formalism. Typically these works have assumed that the Hamiltonian is static. For the more general case of explicitly time dependent operators which we wish to consider, certain simplifying features of the static case are absent. In these circumstances it is found that the backward evolution equation is the appropriate Heisenberg equation of motion. Subsection 4a explains this and summarizes the properties of evolution operator $\Gamma_{s,t}$. In the next subsection the alternative method is developed and consistency with the results of §3 is considered.

4a. *Heisenberg and Moyal Equations*

The essential features of the various possible Heisenberg equations of motion which need to be clarified are all present at the operator level, so we set this discussion there.

Let $\hat{A}(t)$, and $\hat{H}(t)$, be a time dependent Schrödinger picture operator, and Hamiltonian. In the Heisenberg picture,

$$\hat{A}(t,s) = \boldsymbol{\Gamma}(s,t)\hat{A}(t) = U(t,s)^\dagger \, \hat{A}(t) \, U(t,s), \tag{4.1}$$

and $\hat{H}(t,s) = \boldsymbol{\Gamma}(s,t)\hat{H}(t)$. Note that if $\hat{H}(t)$ is not static, then $U(t,s)$ is not $e^{(t-s)\hat{H}/i\hbar}$ and it need not commute with $\hat{H}(t)$, so $\hat{H}(t,s) \neq \hat{H}(t)$. The 'forward' and 'backward' Heisenberg equations of motion are, respectively,

$$\frac{\partial}{\partial t}\hat{A}(t,s) = \frac{1}{i\hbar}\big[\hat{A}(t,s)\,,\,\hat{H}(t,s)\big] + \frac{d\hat{A}}{dt}(t,s), \tag{4.2a}$$

$$\frac{\partial}{\partial s}\hat{A}(t,s) = -\frac{1}{i\hbar}\big[\hat{A}(t,s)\,,\,\hat{H}(s)\big], \tag{4.2b}$$

where $\frac{d\hat{A}}{dt}(t,s) \equiv \boldsymbol{\Gamma}(s,t)\big(d\hat{A}(t)/dt\big)$. These follow from the four variants of the Schrödinger equation (1.3) in which either $U$ or $U^\dagger$ is differentiated with respect to either its first or second argument (in deriving these, $U(t,s)^\dagger = U(t,s)^{-1} = U(s,t)$ is used).

The following features of the Heisenberg equations are critical. The backward equation (4.2b) involves the commutator with the Schrödinger picture Hamiltonian $\hat{H}(s)$. It is clearly a homogeneous linear non-autonomous evolution equation for $\hat{A}(t,s)$ with $s$ playing the role of the evolution parameter. Time $t$ is passive, and forms a convenient 'pinning time' for the initial condition

$$\hat{A}(t,\tau)\big|_{\tau=t} = \hat{A}(t). \tag{4.3}$$

By contrast the forward equation is *not* a homogeneous evolution equation for $\hat{A}(t,s)$: the Schrödinger operator $d\hat{A}(t)/dt$ is an arbitrary prescribed function used to form an inhomogeneous term. Moreover, the commutation operator $(i\hbar)^{-1}\big[\,\cdot\,,\,\hat{H}(t,s)\big]$ involves the Heisenberg picture Hamiltonian. Therefore both coefficients $\hat{H}(t,s)$ and $\frac{d\hat{A}}{dt}(t,s)$ are unknown (since $U(t,s)$ is), and their dependence on the initial time $s$ violates the standard formulation of an evolution equation. A proper evolution equation is, however, obtained from (4.2a) in the special case where $\hat{A}$ and $\hat{H}$ are static.

Thus, it is the backward Heisenberg equation which provides a solvable statement of the evolution problem in the general case of time dependent operators. An intuitive understanding of this fact is also useful. Equation (4.1) shows that $\boldsymbol{\Gamma}(s,t)$ maps an object 'at' time $t$, namely $\hat{A}(t)$, into one at time $s$ (recall that $\hat{A}(t,s)$ is traced with states frozen at time $s$ to yield expectation values). In this sense then, the initial time parameter of the Schrödinger equation is in fact the natural forward evolution parameter for the Heisenberg picture. We now denote it by $\tau$, as was done in (1.1). For other applications of the backward Heisenberg equations see Refs. [44,45].

The Wigner–Weyl image of the preferred Heisenberg equation (4.2b) is the Moyal equation (1.1). Introducing $\Gamma_{\tau,t}$ as in (1.7) it reads

$$\frac{\partial}{\partial \tau}\Gamma_{\tau,t}(a) = -\left\{\Gamma_{\tau,t}(a)\,,\,H_{\mathrm{w}}(\tau)\right\}_{\mathrm{M}}, \tag{4.4a}$$

for arbitrary phase space functions $a$. The companion initial condition (4.3) becomes

$$\Gamma_{t,t}(a) = a. \tag{4.4b}$$

(Identity (4.4a) appeared in the symbolic form (1.12).)

Note that (4.4a) contains the Moyal bracket rather than the Poisson bracket. Roughly, this means it involves infinite order differentiation of $\Gamma_{\tau,t}(a)$ [cf. (2.18)] and so it is not a first order transport equation like (1.2). A rigorous analysis[23] would interpret $\left\{\,\cdot\,,\,H_{\mathrm{w}}(\tau)\right\}_{\mathrm{M}}$ as a pseudodifferential operator acting in a Banach space of functions, say $\mathcal{F} \ni a$. Under suitable conditions[46] on this operator, (4.4) will be a linear evolution problem with a unique solution defined for $\tau \in \mathbb{R}$. We shall proceed under the assumption that this is the case. A parallel situation exists regarding the operator valued Schrödinger equation (1.3). Restrictive assumptions on $\hat{H}(\tau)$ must be imposed in order to ensure the applicability of a theory of linear evolution equations in $\mathcal{H}$ in a form that guarantees the existence of a unitary solution. For example, if the self-adjoint family of operators $\hat{H}(\tau)$ has a stable domain and on this domain is strongly continuously differentiable in $\tau$ then this suffices to prove (e.g. Refs. [47,48]) the existence of a unique $U(\tau,s)$.

Clearly, $\Gamma_{s,t}$ is a kind of fundamental solution for the Moyal equation. As the next lemma shows, it possesses all the properties characteristic of an evolution family.

**Lemma 4.** (Properties of $\Gamma_{s,t}$)   For all $s,t \in \mathbb{R}$
(a) $\Gamma_{s,t} : \mathcal{F} \to \mathcal{F}$ is $\mathbb{C}$-linear.
(b) $\Gamma_{s,t}$ is bijective and $\Gamma_{s,t}^{-1} = \Gamma_{t,s}$.
(c) An evolution law is obeyed: for all $\rho \in \mathbb{R}$

$$\Gamma_{s,t} = \Gamma_{s,\rho} \circ \Gamma_{\rho,t}. \tag{4.5}$$

(d) $\Gamma_{t,t} = \mathrm{Id}_{\mathcal{F}}$
(e) If $1 \in \mathcal{F}$ is the constant function with value 1, then $\Gamma_{s,t}1 = 1$.
(f) $\Gamma_{s,t}$ preserves the $*$-product: for all $a,b \in \mathcal{F}$

$$\Gamma_{s,t}(a) * \Gamma_{s,t}(b) = \Gamma_{s,t}(a * b). \tag{4.6}$$

Hence $\Gamma_{s,t}$ preserves the Moyal bracket too.

If $\Gamma_{s,t}$ is defined by (1.7) then the proofs of these properties follow routinely from properties of $U(t,s)$. Alternatively, one can define $\Gamma_{s,t}(a)$ as the unique solution of (4.4).

The proof of Lemma 4 then relies in a central way on this unique solvability hypothesis, and is not difficult so we omit it. The index order chosen for $\Gamma_{s,t}$ is in harmony with evolution law (4.5). Taken together, properties (a), (b) and (f) state that each $\Gamma_{s,t}$ is an automorphism of the algebra $\mathcal{F}$ with product $*$.

The above properties of $\Gamma_{s,t}$ do not display any feature which is parallel to the unitarity of the Schrödinger evolution $U(t,s)$. It is, however, possible to formulate the problem in such a way that $\Gamma_{s,t}$ is unitary. Let $\mathcal{B}_2$ denote the Schmidt class of operators[49] acting in $\mathcal{H} = L^2(\mathbb{R}^d)$. It is known[50] that $\mathcal{B}_2$ is a Hilbert space with respect to the inner product $(\hat{A}, \hat{B}) \mapsto \mathrm{Tr}(\hat{A}^\dagger \hat{B})$. From the unitarity of $U(t,s)$ it is simple to check that the Heisenberg evolution $\boldsymbol{\Gamma}(s,t)$ — acting in $\mathcal{B}_2$ — is unitary. Now, the Wigner transform $\sigma_\mathrm{w}$ is a Hilbert space isomorphism from $\mathcal{B}_2$ onto $\mathcal{F} \equiv \mathcal{L}^2(\mathbb{R}^{2d}, h^{-d} d\zeta)$ [cf. (5.2a)]. Thus, it follows that the Heisenberg–Weyl evolution map $\Gamma_{s,t} = \sigma_\mathrm{w} \circ \boldsymbol{\Gamma}(s,t) \circ \sigma_\mathrm{w}^{-1}$ is unitary in this space $\mathcal{F}$ of square integrable symbols.

An interesting application of the properties above is to verify preservation of the canonical Moyal brackets (1.10).

**Corollary.** Let $Z_\mathrm{w}^i(t,s) \equiv \Gamma_{s,t}(\pi^i)$, where $\pi^i$ are Euclidean coordinates on $\mathbb{R}^{2d}$, i.e. $\pi^i = \sigma_\mathrm{w}(\hat{z}^i)$ is projection onto the $i^\mathrm{th}$ component. Then (1.10) holds.

**Proof:** Since $\pi^i$ supports only one nonzero derivative, $\{\pi^i, \pi^j\}_\mathrm{M} = \{\pi^i, \pi^j\} = J^{ij}$. From this and properties (a), (e) and (f) one computes

$$\{Z_\mathrm{w}^i(t,s), Z_\mathrm{w}^j(t,s)\}_\mathrm{M} = \Gamma_{s,t}\left(\{\pi^i, \pi^j\}_\mathrm{M}\right) = \Gamma_{s,t}(J^{ij}) = J^{ij}.$$

∎

An advantage of the Weyl symbol representation is that it is possible to identify a natural $\hbar$-parity of quantum mechanics. Commonly occurring physical systems have Weyl symbols (3.1) that are finite order in $\hbar$ with all $\hbar$-odd terms absent. In this standard situation $H_\mathrm{w}(\tau, \hbar; \zeta)$ has even $\hbar$-parity and, as the lemma below shows, so does the Heisenberg–Weyl evolution operator $\Gamma_{s,t}$. This simplifies the semiclassical expansion (1.8) since all $\gamma_{s,t}^{(n)}$ with odd $n$ will vanish.

**Lemma 5.** (Evolution $\hbar$-Parity) Assume the Hamiltonian's Weyl symbol $H_\mathrm{w}(\tau, \hbar; \cdot)$ has even dependence on $\hbar$. Then $\Gamma_{s,t}$ is even in $\hbar$, and so all $\gamma_{s,t}^{(2n+1)} = 0$.

**Proof:** The map $\Gamma_{s,t}^\hbar$ (where the $\hbar$ dependence has been made explicit) is defined as the unique solution to (4.4). If $\hbar$ is replaced by $-\hbar$ throughout these equations, it is found that $\Gamma_{s,t}^{-\hbar}$ also obeys (4.4) because $\{\cdot, \cdot\}_\mathrm{M}$ is $\hbar$-even (cf. (2.18)) and $H_\mathrm{w}(\tau)$ is assumed to be so. By uniqueness of the solution, $\Gamma_{s,t}^\hbar = \Gamma_{s,t}^{-\hbar}$. ∎

*4b. Classical Inhomogeneous Transport Equations*

Consider the Heisenberg–Moyal evolution problem (4.4). The Hamiltonian symbol $H_w(\tau)$ is given by (3.1). We take (1.8) as an ansatz for the analytic $\hbar$ structure of the Heisenberg–Weyl evolution map $\Gamma_{s,t}$. The initial conditions on $\gamma_{s,t}^{(n)}$ induced from (4.4b) are

$$\gamma_{t,t}^{(0)} = \mathrm{Id}_{\mathcal{F}}, \qquad \gamma_{t,t}^{(n)} = 0 \quad (n \geq 1). \tag{4.7}$$

If these two representations are substituted into (4.4a) and the Moyal bracket $\hbar$ expansion (2.18) is used, then an $\hbar$-expanded version of the Moyal equation results. Equating in it the coefficients of $\hbar^r$, $r \geq 0$, leads to the following identities which

$$\mathbf{a}_r(t, s; \,\cdot\,) \equiv \gamma_{s,t}^{(r)}(a) \tag{4.8}$$

must satisfy:

$$\partial_2 \mathbf{a}_r(t, \tau; \zeta) + \{\mathbf{a}_r(t, \tau; \,\cdot\,), H_c(\tau, \,\cdot\,)\}(\zeta) = f_r(t, \tau, \zeta), \tag{4.9a}$$

$$f_r(t, \tau, \zeta) \equiv \sum_{\substack{j+k+2l=r \\ k<r}} \frac{(-)^{l+1} r!}{2^{2l}(2l+1)!\, j!\, k!} B^{2l+1} \prec \mathbf{a}_k(t, \tau; \,\cdot\,), h_j(\tau, \,\cdot\,) \succ (\zeta). \tag{4.9b}$$

Here $\partial_2$ is the partial derivative on the second time argument, and the sum is over all integers $j, k, l \geq 0$ obeying the two indicated constraints. By definition $f_0 = 0$. Equation (4.9a) is an inhomogeneous linear Hamiltonian transport equation. For $r = 0$ it reduces to the homogeneous equation (1.2). The inhomogeneous term $f_r$ is built from $\mathbf{a}_0, \ldots, \mathbf{a}_{r-1}$, which leads to recursive solutions for $\gamma_{s,t}^{(r)}$. The unique solvability of (4.9) is based upon the following results.

**Lemma 6.** (Backward Flow) The flow $g(t, s | \,\cdot\,)$ of a time dependent Hamiltonian $H_c(t, \,\cdot\,)$ obeys

$$\partial_2 g(t, \tau | \zeta) = -\nabla g(t, \tau | \zeta) J \nabla H_c(\tau, \zeta). \tag{4.10}$$

**Proof:** Differentiating the flow identity

$$g(\tau, t | g(t, \tau | \zeta)) = \zeta \tag{4.11}$$

with respect to $\tau$ using the chain rule and solving for $\partial_2 g$ yields

$$\partial_2 g(t, \tau | \zeta) = -\nabla g(\tau, t | z)^{-1} \partial_1 g(\tau, t | \zeta),$$

where $z \equiv g(t, \tau | \zeta)$. Now differentiate (4.11) with respect to $\zeta$ to find $\nabla g(\tau, t | z)^{-1} = \nabla g(t, \tau | \zeta)$. Substituting $\nabla g$ from here and $\partial_1 g$ from Hamilton's equation (1.6) results in (4.10). Compare also Lemma 4.2.32 of Ref. [13]. ∎

**Theorem 2.** Let the Hamiltonian $H_c(\tau, \zeta)$, inhomogeneous term $f(\tau, \zeta)$ and initial data $\beta(\zeta)$ be smooth functions, with $g$ the flow of $H_c$. Then the unique smooth solution $b$ of the Hamiltonian transport problem

$$\partial b(\tau, \zeta) + \{b(\tau, \cdot\,), H_c(\tau, \cdot\,)\}(\zeta) = f(\tau, \zeta) \tag{4.12a}$$

$$b(t, \zeta) = \beta(\zeta) \tag{4.12b}$$

is given by

$$b(\tau, \zeta) = \beta\bigl(g(t, \tau|\zeta)\bigr) - \int_\tau^t d\rho\, f\bigl(\rho, g(\rho, \tau|\zeta)\bigr). \tag{4.13}$$

**Proof:** [Uniqueness] Let $b$ be a solution of (4.12). Upon substituting $\zeta = g(\tau, t|z)$ into (4.12a) it follows from Hamilton's equation that the left side assumes the form of a total derivative, so $(d/d\tau)\, b\bigl(\tau, g(\tau, t|z)\bigr) = f\bigl(\tau, g(\tau, t|z)\bigr)$. Integrating this ODE from $\tau$ to $t$ and using (4.12b) yields

$$b\bigl(\tau, g(\tau, t|z)\bigr) = \beta(z) - \int_\tau^t d\rho\, f\bigl(\rho, g(\rho, t|z)\bigr).$$

Finally, choose $z = g(t, \tau|\zeta)$ to find that $b$ must be given by (4.13).

[Existence] If $b$ is defined by (4.13) it clearly obeys (4.12b), so it must be shown to satisfy (4.12a). Differentiating (4.13) with respect to $\tau$ and using (4.10) to represent $\partial_2 g$ leads to

$$\partial b(\tau, \zeta) = \Bigl[ -\nabla\beta\bigl(g(t, \tau|\zeta)\bigr) \cdot \nabla g(t, \tau|\zeta) + \int_\tau^t d\rho\, \nabla f\bigl(\rho, g(\rho, \tau|\zeta)\bigr) \cdot \nabla g(\rho, \tau|\zeta) \Bigr] J \nabla H_c(\tau, \zeta) + f(\tau, \zeta). \tag{4.14}$$

By the chain rule, $\nabla \beta(g) \cdot \nabla g = \partial \beta\bigl(g(t, \tau|\zeta)\bigr)/\partial \zeta$, with a similar reduction for $\nabla f \cdot \nabla g$ in the integral term. The first term on the right of (4.14) thus reduces to a Poisson bracket, specifically $-\{b(\tau, \cdot\,), H_c(\tau, \cdot\,)\}$. Thus $b$ obeys (4.12a). ∎

Application of Theorem 2 to the transport problem (4.9) and (4.7), of quantum origin, immediately yields the solutions

$$\mathbf{a}_r(t, \tau; \zeta) = \delta_{r,0}\, a\bigl(g(t, \tau|\zeta)\bigr) + \sum_{\substack{j+k+2l=r \\ k<r}} \frac{(-)^l r!}{2^{2l}\,(2l+1)!\, j!\, k!} \\ \times \int_\tau^t d\rho\, B^{2l+1} \prec \mathbf{a}_k(t, \rho; \cdot\,),\, h_j(\rho, \cdot\,) \succ \bigl(g(\rho, \tau|\zeta)\bigr). \tag{4.15}$$

For $r = 0$ the integral term is absent and $\mathbf{a}_0(t, \tau; \zeta)$ here agrees with (3.37). For $r \geq 1$, equations (4.15) provide a recursive integral solution based on classical transport.

It is instructive to see the explicit form of the first few coefficients $\mathbf{a}_r$. We consider a system with a Weyl quantized Hamiltonian; accordingly $h_j = 0$. The parity property of $\Gamma_{s,t}$ assures us that all the odd index coefficients $\mathbf{a}_n$ vanish. The predictions of (4.15) for the non-vanishing terms to order $\hbar^6$ are:

$$\mathbf{a}_2(t,s;\zeta) = \int_s^t d\rho \frac{-1}{12} B^3 \prec \mathbf{a}_0(t,\rho;\,\cdot\,), H_c(\rho,\,\cdot\,) \succ (g(\rho,s|\zeta))$$

$$\mathbf{a}_4(t,s;\zeta) = \int_s^t d\rho \left\{ \frac{1}{80} B^5 \prec \mathbf{a}_0(t,\rho;\,\cdot\,), H_c(\rho,\,\cdot\,) \succ - \frac{1}{2} B^3 \prec \mathbf{a}_2(t,\rho;\,\cdot\,), H_c(\rho,\,\cdot\,) \succ \right\} (g(\rho,s|\zeta))$$

$$\mathbf{a}_6(t,s;\zeta) = \int_s^t d\rho \left\{ \frac{-1}{448} B^7 \prec \mathbf{a}_0(t,\rho;\,\cdot\,), H_c(\rho,\,\cdot\,) \succ + \frac{3}{16} B^5 \prec \mathbf{a}_2(t,\rho;\,\cdot\,), H_c(\rho,\,\cdot\,) \succ \right.$$

$$\left. - \frac{5}{4} B^3 \prec \mathbf{a}_4(t,\rho;\,\cdot\,), H_c(\rho,\,\cdot\,) \succ \right\} (g(\rho,s|\zeta)). \tag{4.16}$$

Further simplifications of the recursive representation (4.15) arise if $H_c(\tau,\zeta)$ is a polynomial function of order $N$. In this case $H_c$ can support at most $N$ non-vanishing derivatives, so all factors with $B^{2l+1} \prec \mathbf{a}_k(t,\rho;\,\cdot\,), H_c(\rho,\,\cdot\,) \succ$ are zero if $2l+1 > N$. An important special case of this arises if $H_c(\tau,\zeta)$ is quadratic. As is evident from (4.16), $\mathbf{a}_n = 0$ for $n > 0$, and the zeroth order semiclassical expansion (1.8) is then exact. Specifically, $\Gamma_{s,t} = \gamma_{s,t}^{(0)}$.

The iterated Poisson bracket transport formulae (4.15) and (4.16) are remarkably concise representations. This simplicity is somewhat deceptive due to their recursive nature. The computation of $\mathbf{a}_2$ in terms of the original symbol $a$, for example, requires the substitution of $\mathbf{a}_0 = a \circ g$ into the first of (4.16) with subsequent evaluation of its third derivative. Further iteration to compute $\mathbf{a}_4$, etc., will lead to more derivatives and multiple time integrals

Representations of the Heisenberg–Weyl semiclassical expansion coefficients $\gamma_{s,t}^{(n)}$ are obtained from (4.15) by making the notational change indicated in (4.8). For example, the first correction is

$$\gamma_{s,t}^{(1)} a(\zeta) = \int_s^t d\rho \left\{ a\bigl(g(t,\rho|\,\cdot\,)\bigr),\, h_1(\rho,\,\cdot\,) \right\} (g(\rho,s|\zeta)) \tag{4.17a}$$

$$= \left\{ a\bigl(g(t,s|\,\cdot\,)\bigr),\, \int_s^t d\rho\, h_1\bigl(\rho, g(\rho,s|\,\cdot\,)\bigr) \right\}(\zeta), \tag{4.17b}$$

where invariance of the Poisson bracket under the symplectic map $g$ was used. Similarly, equations (4.16) and (4.8) immediately provide recursive forms of $\gamma_{s,t}^{(2)}$, $\gamma_{s,t}^{(4)}$ and $\gamma_{s,t}^{(6)}$ in the case of a Weyl quantized Hamiltonian.

We close this subsection by verifying that representation (4.17) of $\gamma_{s,t}^{(1)}$ agrees with the previous result (3.38). To do this, simply substitute (3.28) into (3.38) and note that the derivative structure $\nabla_i a(g) \nabla_j g_c^i = \nabla_j (a \circ g)$, again by the chain rule. In this way (4.17b) emerges.

## V — SEMICLASSICAL BEHAVIOR OF EXPECTATION VALUES

The preceding sections have developed, in considerable detail, methods for computing the small $\hbar$ expansion of Heisenberg picture operators. From a physical standpoint it is ultimately desirable to employ these results in the broader context of observable dynamical phenomena. In particular, a semiclassical treatment of quantum expectation values is called for. Viewed in this way, the $\hbar$ expansion of $A_{\rm w}(t,s)$ forms one individual theme, which should be combined with an appropriate description of quantum states to yield a semiclassical analysis of the expectation value. In this section we consider these questions in a general way. Our objective here is not to construct detailed higher order expansion formulae, but rather it is twofold: (1) to sketch in rough outline the nature and advantages of the theory of expectation values in the Heisenberg picture and Weyl representation; (2) to indicate the evident physical character of the results which can be obtained.

In general a quantum state is a density operator $\hat{\rho}$, that is, a positive trace-class operator of unit trace, $\operatorname{Tr}\hat{\rho} = 1$. The expectation value of observable $\hat{A}$ in the state $\hat{\rho}$ is $\langle \hat{A} \rangle_{\hat{\rho}} = \operatorname{Tr}(\hat{A}\hat{\rho})$. The Weyl symbol versions of these concepts are well known. The trace becomes a phase space integral

$$\operatorname{Tr} \hat{A} = h^{-d} \int_{\mathbb{R}^{2d}} d\zeta \, A_{\rm w}(\zeta), \tag{5.1}$$

as follows at once from (2.6). To obtain the trace of a product one can combine Groenewold's formula (2.9) with the identity

$$\int d\zeta \, f(\nabla_1 \cdot J \nabla_2) \prec A_{\rm w}, B_{\rm w} \succ (\zeta) = \int d\zeta \, f(0) A_{\rm w}(\zeta) B_{\rm w}(\zeta)$$

(which follows for analytic $f$ by formally integrating $\nabla_2$ by parts); thus one finds

$$\operatorname{Tr}\left(\hat{A}(t,s)\hat{\rho}\right) = h^{-d} \int d\zeta \, A_{\rm w}(t,s;\zeta) \rho_{\rm w}(\zeta). \tag{5.2a}$$

We summarize a few elementary analytical properties[2,3,51-54] of densities. The symbol $\rho_{\rm w}$ is a real valued, uniformly continuous $L^2$ function on $\mathbb{R}^{2d}$, which vanishes at infinity and is bounded by $|\rho_{\rm w}| \leq (2/h)^d$. The integral $h^{-d} \int d\zeta \, \rho_{\rm w}(\zeta) = 1$ corresponds to the unit trace normalization, via (5.1). Unlike a classical probability density, the quantum density $\rho_{\rm w}$ can take on negative values.

An important class of states are the pure states. These are 1-dimensional projection operators. If $\psi \in \mathcal{H}$ is a normalized vector in the range of such a pure state $\hat{\rho}$, then $\hat{\rho} = |\psi\rangle\langle\psi|$ and (5.2a) becomes

$$\langle \hat{A}(t,s) \rangle_{\psi} = \int d\zeta \, A_{\rm w}(t,s;\zeta) w_{\psi}(\zeta), \tag{5.2b}$$

where $w_\psi \equiv \hbar^{-d}|\psi\rangle\langle\psi|_{\text{w}}$ is the 'Wigner function' of $\psi$.

Of critical importance to semiclassical analysis is an understanding of the $\hbar$ analytic structure of pure state densities. The projection property of pure states, $\hat{\rho}^2 = \hat{\rho}$, in symbol form reads $\rho_{\text{w}} * \rho_{\text{w}} = \rho_{\text{w}}$, or using Groenewold's formula,

$$e^{(i\hbar/2)B_{12}} \prec \rho_{\text{w}}, \rho_{\text{w}} \succ = \rho_{\text{w}}. \tag{5.3}$$

Suppose, for example, that one may represent the structure of $\rho_{\text{w}}$ near $\hbar = 0$ by a Laurent series,

$$\rho_{\text{w}}(\hbar; \zeta) = \sum_{j=M}^{\infty} \hbar^j \rho_j(\zeta) \qquad (\rho_M \neq 0),$$

where either $M \in \mathbb{Z}$ (the integers) or $M = -\infty$, and substitute this series into (5.3) and match common powers of $\hbar$. It will be found that either $M = 0$ or $M = -\infty$. The only regular ($M = 0$) series consistent with (5.3) is $\rho_{\text{w}} = 1$, which is not a density (it is not normalizable). Thus $M = -\infty$ and we conclude that the Wigner function of a pure state $w_\psi$ (or a finite linear combination of pure states) cannot be regular at $\hbar = 0$, and if it has an isolated singularity there, it must be an essential singularity. Note that this conclusion holds regardless of whether or not the wave function $\psi(x)$ representing the pure state has explicit $\hbar$ dependence.

By contrast, mixed states ($\hat{\rho}^2 \neq \hat{\rho}$) can have $\hbar$-regular densities. An example is the density $\hat{\rho}(\beta) \equiv e^{-\beta\hat{H}}/\text{Tr}\, e^{-\beta\hat{H}}$ of the canonical ensemble at inverse temperature $\beta$. In this case the $\hbar$-regular expansion of $\hat{\rho}(\beta)_{\text{w}}$ is the venerable Wigner–Kirkwood[2,55−58] expansion. This expansion can be derived by proceeding as in §2d after replacing $t \to \beta\hbar/i$.

Note that equations (5.2) have been written in the Heisenberg picture, where the time evolution is ascribed to the observable $\hat{A}(t, s)$ and the state, $\hat{\rho}$ or $\psi$, is held fixed. This is in keeping with the expansions (3.33) and (4.15). Much of the prior literature[39,46,59] uses the Schrödinger picture in which the density (initially $\hat{\rho}$) carries the time dependence according to $\hat{\rho}(t, s) = U(t, s)\hat{\rho}\, U(t, s)^\dagger = \boldsymbol{\Gamma}(t, s)\hat{\rho}$. Here the roles of $t$ and $s$ are opposite to the observable case (1.4), and $\hat{\rho}(t, s)$ obeys the quantum Liouville equation of motion

$$\frac{\partial}{\partial t}\hat{\rho}(t, s) = \frac{1}{i\hbar}\big[\hat{H}(t), \hat{\rho}(t, s)\big] \tag{5.4}$$

(usually considered for static Hamiltonians where $\hat{H}(t) = \hat{H}$).

Clearly, one could try to apply our results for the Heisenberg–Weyl evolution map to compute $\rho_{\text{w}}(t, s) = \Gamma_{t,s}(\rho_{\text{w}})$. However, this Schrödinger picture approach encounters considerable inherent difficulties due to the fact, noted above, that typically the initial density $\rho_{\text{w}}$ is essentially singular in $\hbar$. Clearly one would have to determine the precise nature of the essential singularity as it evolves in time, and develop techniques to construct the small $\hbar$ perturbative form of the residual nonsingular factor, if one is to get useful semiclassical expansion formulae. This is one of the reasons we are led to the Heisenberg picture, where the observable maintains a regular (semiclassically admissible) $\hbar$ analytic form.

In the circumstances where the density $\rho_w$ carries an $\hbar$ essential singularity, the determination of the semiclassical form of the expectation value $\langle \hat{A}(t,s) \rangle_\psi$ requires a careful evaluation of the phase space integral (5.2$b$). We investigate this topic, in the preferred Heisenberg picture, by considering several informative examples.

**Example 1.** (Semiclassical States and Interference)   Consider a state vector $\psi$ of semiclassical form in the configuration representation,

$$\psi(x) = \varphi(x) \exp\{\frac{i}{\hbar} S(x)\}, \tag{5.5}$$

with smooth amplitude $\varphi \in L^2(\mathbb{R}^d)$ and smooth real phase function $S$. Combining (5.2$b$) and (2.7), the Wigner function based expectation of observable $\hat{A}(t,s)$ in this state is

$$\langle \hat{A}(t,s) \rangle_\psi = \int dq\, (2\pi\hbar)^{-d} \int\int dpdv\, A_w(t,s,\hbar;q,p) \varphi(q+v/2)\bar\varphi(q-v/2) \\ \times \exp\{\frac{i}{\hbar}[-p\cdot v + S(q+v/2) - S(q-v/2)]\}. \tag{5.6}$$

For each fixed $q$, we will find the leading $\hbar \downarrow 0$ asymptotics of the $(p,v)$-integral in (5.6) by employing the stationary phase asymptotic formula [Ref. [60], Theorem 2.2]

$$(2\pi\hbar)^{-n/2} \int d^n y\, F(y,\hbar) e^{(i/\hbar)\Phi(y)} \tag{5.7}$$
$$= \left|\det \nabla\nabla\Phi(y_o)\right|^{-1/2} F(y_o,\hbar) e^{(i/\hbar)\Phi(y_o)} e^{i(\pi/4)\mathrm{sign}\,\nabla\nabla\Phi(y_o)} + \mathcal{O}(\hbar^{1/2}).$$

In this formula $y_o$ denotes the critical point $(\nabla\Phi(y_o)=0)$ of the smooth phase function $\Phi$, whose Hessian is $\nabla\nabla\Phi$. In applying (5.7) to (5.6), identify $y=(p,v)$, hence $n=2d$, and $\Phi(p,v) = -p\cdot v + S(q+v/2) - S(q-v/2)$, while $F(p,v,\hbar)$ is the non-exponential portion of the integrand. The unique critical point of $\Phi$ is easily found to be given by

$$p_o = \nabla S(q), \qquad v_o = 0, \tag{5.8}$$

while the Hessian there is

$$\nabla\nabla\Phi(p_o,v_o) = -\begin{bmatrix} 0 & \delta \\ \delta & 0 \end{bmatrix}. \tag{5.9}$$

It is readily demonstrated that this matrix has the eigenvalues $+1$ and $-1$, each of multiplicity $d$. Hence the signature is zero. Concerning $A_w(t,s,\hbar;q,p)$, we will keep its leading semiclassical value $\mathbf{a}_0(t,s;q,p)$ as given by (3.34). Employing (5.7)–(5.9) in this way gives

$$\langle \hat{A}(t,s) \rangle_\psi = \int dq\, A_c\bigl(t, g(t,s|q,\nabla S(q))\bigr)\, |\psi(q)|^2 + \mathcal{O}(\hbar^{1/2}). \tag{5.10}$$

Formula (5.10) has a transparent physical interpretation, insofar as $|\psi(q)|^2$ is the probability density at $q$ and $\nabla S(q)$ is the leading order momentum of $\psi$. Note that while the Wigner function $w_\psi$ has an essential singularity in $\hbar$, nevertheless once the momentum integration is completed the result is an $\hbar$-regular expectation value.

It is of interest to generalize this example by considering a state which is a (normalized) superposition of two semiclassical wave packets like (5.5),

$$\psi = \psi_1 + \psi_2 = \varphi_1 e^{iS_1/\hbar} + \varphi_2 e^{iS_2/\hbar}.$$

The corresponding density

$$\hat{\rho} = \hat{\rho}_1 + \hat{\rho}_2 + |\psi_1\rangle\langle\psi_2| + |\psi_2\rangle\langle\psi_1|$$

is still a pure state, $|\psi\rangle\langle\psi|$, but the cross terms will now describe interference of the two packets. In computing $\langle \hat{A}(t,s) \rangle_\psi$ the terms $\hat{\rho}_1$ and $\hat{\rho}_2$ will each give a contribution of the form (5.10). The stationary phase analysis in (5.6)–(5.10) readily extends to the cross term integrals. The rapidly oscillating phase is now $\Phi(p,v) = -p \cdot v + S_1(q+v/2) - S_2(q-v/2)$, with critical point $v_o = 0$ and $p_o = \frac{1}{2}\nabla(S_1 + S_2)(q)$. Again the Hessian $\nabla\nabla\Phi(p_o, v_o)$ has determinant $(-)^d$ and signature 0. To within $\mathcal{O}(\hbar^{1/2})$ the cross term contribution is found to be

$$\langle \hat{A}(t,s) \rangle_{\text{cross}} = \int dq\, A_c\Big(t, g\big(t,s|q, \tfrac{1}{2}\nabla(S_1 + S_2)(q)\big)\Big)\, 2\text{Re}[\varphi_1 \bar{\varphi}_2 e^{i(S_1-S_2)/\hbar}](q). \quad (5.11)$$

The surviving essential singularity, with a phase difference $S_1 - S_2$, is characteristic of interference phenomena. Formula (5.11) is suitable for computing the effects of wave packet interference at the physical value of $\hbar$. It should be kept in mind that $\varphi_1$ and $\varphi_2$ are initial, not evolving, amplitudes. Of course, if $\hbar \downarrow 0$ this term oscillates to zero whenever $S_1 \neq S_2$.

The computation of higher order $\hbar$ corrections to the leading terms, (5.10) and (5.11), would require a more detailed asymptotic analysis of the Wigner function and the phase space integral. This is beyond our scope here, which is to indicate the character of the obtainable result. A more comprehensive investigation of the $\hbar$ structure of physically relevant quantum densities $\rho_w$ is a significant topic in its own right, and is the subject of ongoing work[38,61,62] in the literature. The fact that in (5.10) and (5.11) the formal phase space symmetry between $q$ and $p$ is broken is merely the result of employing a configuration space wave function $\psi(x)$ in these examples.

**Example 2.** (Harmonic Oscillator)    Consider the 1-dimensional harmonic oscillator with unit mass, $\hat{H} = (\hat{p}^2 + \omega^2 \hat{q}^2)/2$, and choose the initial state to be the $\hbar$ independent Gaussian

$$\phi(x) = c\, e^{-x^2/2}, \qquad c = \pi^{-1/4}. \quad (5.12)$$

For illustrative purposes we pick the observable $\hat{A} = \hat{q}^2$, and shall compare the computation of $\langle \hat{q}^2 \rangle_{\psi_t}$, $\psi_t = U(t,0)\phi$, via two methods: (i) standard propagator based quantum mechanics, and (ii) the Heisenberg–Weyl formalism. Because of the simple quadratic Hamiltonian, both of these approaches yield closed form solutions.

**(i)** The lowest order WKB approximation[63–68] gives exact representations for the propagators of $(\hat{q}, \hat{p})$-quadratic Hamiltonians. For the harmonic oscillator the resulting expression is known as Mehler's formula[69,70]

$$K(t,x,y) \equiv \langle x | U(t,0) | y \rangle = \left| \frac{\omega/\hbar}{\sin \omega t} \right|^{\frac{1}{2}} e^{-i(\pi/4)\mu(t)} \exp\left\{ \frac{i\omega}{2\hbar}[(x^2 + y^2)\cot \omega t - 2xy \csc \omega t] \right\}, \tag{5.13a}$$

valid for $t \neq n\pi/\omega$ ($n \in \mathbb{Z}$), with the Maslov index $\mu$ being constant between successive conjugate points: $\mu(t) = 2n + 1$ for $n\pi < \omega t < (n+1)\pi$. At a caustic the propagator is a distribution,

$$K\left(\frac{n\pi}{\omega}, x, y\right) = e^{-in\pi/2}\, \delta\!\left((-)^n x - y\right). \tag{5.13b}$$

The fact that $\mu(t)$ varies discontinuously in going through a caustic is a necessary feature of the propagator, which ensures that $K$ is the kernel of the $t$-strongly continuous evolution operator $U(t,0)$. The fact that $\mu$ depends only on $t$ is a special feature of static quadratic Hamiltonians, such as the harmonic oscillator.

Calculating for $t \neq n\pi/\omega$, the evolved state is obtained as a complex Gaussian integral

$$\psi_t(x) = \int dy\, K(t,x,y)\phi(y)$$

$$= c \left| \frac{\lambda}{\sin \omega t} \right|^{\frac{1}{2}} e^{-i(\pi/4)\mu(t)} (1 - i\lambda \cot \omega t)^{-1/2} \exp\left\{ \frac{x^2}{2}\left[ i\lambda \cot \omega t - \frac{\lambda^2 \csc^2 \omega t}{1 - i\lambda \cot \omega t} \right] \right\},$$

where $\lambda = \omega/\hbar > 0$ and the branch cut for $(\cdot)^{-1/2}$ is along the negative real axis with argument in $(-\pi, \pi)$. A careful check will show that $\psi_t(x)$ has a removable singularity at $\omega t = n\pi$ and is indeed continuous there.

Once $\psi_t$ is used to calculate an expectation value, $\langle \hat{A} \rangle_{\psi_t} = \int dx\, \bar{\psi}_t(x)[\hat{A}\psi_t](x)$, any effect of the Maslov index will be lost. We take the example $\hat{A} = \hat{q}^2$ and find after a Gaussian integration that

$$\langle \hat{q}^2 \rangle_{\psi_t} = \frac{1}{2}\left( \frac{\hbar^2}{\omega^2} \sin^2 \omega t + \cos^2 \omega t \right). \tag{5.14}$$

Initially this result holds away from caustics, but a separate calculation for $t = n\pi/\omega$ using (5.13b) establishes (5.14) for all $t \in \mathbb{R}$. Note that the final result is regular in $\hbar$ and $t$.

**(ii)** For initial state $\phi$ of (5.12) the Wigner function is simply computed as the Fourier transform of a real Gaussian,

$$w_\phi(q,p) = \frac{c^2}{h} \int_\mathbb{R} dv\, e^{-ipv/\hbar} e^{-(q^2 + v^2/4)} = \frac{1}{\pi \hbar} e^{-(q^2 + p^2/\hbar^2)},$$

and possesses an essential singularity at $\hbar = 0$. This must be paired with the evolved symbol $A_{\mathrm{w}}(t,\hbar;q,p)$. As was shown in §4b, the quadratic nature of $\hat{H}$ implies that $\Gamma_{0,t}$ reduces to its classical part $\gamma_{0,t}^{(0)} = g(t|\,\cdot\,)^*$. Computation of the classical harmonic oscillator flow $g = (q_c, p_c)$ is trivial; one finds $q_c(t|q,p) = (p/\omega)\sin\omega t + q\cos\omega t$. The expectation value is then

$$\langle\, \hat{q}(t,0)^2 \,\rangle_\phi = \int\int dq\,dp\, q_c(t|q,p)^2\, w_\phi(q,p) = \frac{1}{2}\Big(\frac{\hbar^2}{\omega^2}\sin^2\omega t + \cos^2\omega t\Big),$$

in agreement with (5.14).

The examples above, and others[43] from the literature, point to three viable computational methods for extracting the semiclassical behavior of an observable's expectation value. i) Calculate (if possible) the phase space integral (5.2) exactly. This was the approach used in Example 2. ii) Use the stationary phase technique applied to the $(p,v)$ integration in (5.6), in order to obtain the leading small $\hbar$ form of $\langle\, \hat{A}(t,s) \,\rangle_\psi$. This was done in Example 1. iii) Employ a hybrid evaluation of the expectation value. Specifically, in (5.2) keep the exact form of the initial density matrix $\rho_{\mathrm{w}}(\zeta)$ intact with $\hbar$ fixed at its physical value, but replace $A_{\mathrm{w}}(t,s;\zeta)$ with its semiclassical expansion (3.31), truncated at some finite order in $\hbar$ — most often after the leading term $A_c(t,g(t,s|\zeta))$. This approach has the merit of being easy to implement in practical applications. At the same time it respects those aspects of the uncertainty principle (at the physical value of $\hbar$) which are encoded into the initial quantum state.

In a sense, the third approach appears to treat the $\hbar$ dependence in $A_{\mathrm{w}}(t,s;\zeta)$ differently from that in $\rho_{\mathrm{w}}(\zeta)$, but it is a consistent procedure since $A_{\mathrm{w}}(t,s;\zeta)$ is $\hbar$-regular, while $\rho_{\mathrm{w}}(\zeta)$ is typically a localized uniformly bounded $L^1(\mathbb{R}^{2d})$ function. The mathematically distinct asymptotics problem implicit in the hybrid method may at first sight appear unnatural, but it does have a sound physical basis. The limit conventionally described by '$\hbar \downarrow 0$' is physically achieved when some ratio $\hbar^n/S_n$ becomes small, where $S_n$ is some quantity characteristic of the system. Since $\hbar$ is fixed at its physical value, $S_n$ must become large, e.g. by varying the observable or Hamiltonian. By doing this while leaving the initial state fixed, one realizes the asymptotics inherent in the hybrid approach.

Further insight into the Moyal semiclassical method results from a comparison with the WKB approximation,[35,70−72] which provides a semiclassical description of the propagator $K$. The propagator argument $Q \equiv (t,s,x,y)$ defines a set of distinct classical trajectories $\{z_j(\,\cdot\,;Q)\}$ obeying $q_j(s) = y$ and $q_j(t) = x$. Depending on $Q$, the solution set to this two point boundary value problem may be empty, finite or infinite. Locally the trajectories depend continuously on $Q$, but globally $Q$-space is fragmented by surfaces where the nature of the solutions changes. The WKB propagator ansatz[70−72] is a 'sum over paths' of contributions from each trajectory $z_j$ of the form

$$K_j(Q) = (2\pi\hbar)^{-d/2} |D_j(Q)|^{1/2} e^{-i(\pi/4)\mu_j(Q)} \exp\Big\{\frac{i}{\hbar} S_j(Q) + \mathcal{O}(\hbar^1)\Big\},$$

where $S_j$, $D_j$ and $\mu_j$ are, respectively, the principal function (action), its Van Vleck determinant and a Maslov-related phase for the $j^{\text{th}}$ trajectory. The solvability analysis of the Hamiltonian two point boundary value problem and the determination of the Maslov index changes which accumulate through caustics along a trajectory (where $D_j$ becomes singular) are, for general Hamiltonians, significant mathematical problems. Their solution easily becomes prohibitively difficult for practical calculations.

In contrast to this situation, the Heisenberg–Weyl semiclassical method is based upon the classical flow, i.e. the unique classical trajectories which exist for each initial value $\zeta \in \mathbb{R}^{2d}$. Furthermore, there is no need to compute Maslov indices for these trajectories. In summary, it would seem that for many purposes, while the $\langle\, x\,|\, U(t,s)\,|\, y\,\rangle$ based computations are correct, the use of this intermediate tool can introduce significant unnecessary complications due to caustics.

## VI — Conclusion

The Heisenberg picture evolution operator $\boldsymbol{\Gamma}(s,t)$, and equivalently its Weyl phase space counter part $\Gamma_{s,t}$, constitute a complete description of quantum evolution. The Heisenberg–Weyl operator $\Gamma_{s,t}$ — mapping symbols of observables into symbols — admits a regular small $\hbar$ expansion that provides a fundamental statement of semiclassical dynamics.

Explicit forms of the operator valued expansion coefficients $\gamma_{s,t}^{(n)}$ are found by a two stage process. First, the dynamical phase space coordinate operators $\hat{Z}(t,s) \equiv \boldsymbol{\Gamma}(s,t)\hat{z}$, via their symbol $Z_{\text{w}}(t,s,\hbar;\zeta)$, are obtained by using cluster-graph (Theorem 1) and Green function (Lemma 3) methods to formally solve their equation of motion (1.9). The $\hbar$ expansion of $Z_{\text{w}}(t,s,\hbar;\zeta)$ has as its leading term the exact classical flow $g(t,s|\zeta)$ with higher order $\hbar$ terms accounting for quantum fluctuations about the classical motion. Having computed the quantum trajectory $Z_{\text{w}}(t,s,\hbar;\zeta)$ the next stage is to determine the evolution of an arbitrary time dependent observable. Again an application of the cluster expansion yields an exact representation, (3.33), of the dynamically varying observable. Parsing this representation according to powers of $\hbar$ constructs the semiclassical expansion operators $\gamma_{s,t}^{(n)}$.

As the explicit expressions (3.37)–(3.39) for $\gamma_{s,t}^{(n)}$ show, the differential structure of these operators is a universal function of the Poisson bracket operators $B_{ij}$ and finite order phase space gradients. The effect of a specific choice of Hamiltonian $H_{\text{w}}$ is confined to the classical flow arguments $g(t,s|\zeta)$ and the value of the higher order trajectory fluctuations $\mathbf{z}_r(t,s;\zeta)$. For all $\hbar$-even Hamiltonians the odd coefficients $\gamma_{s,t}^{(2n+1)}$ vanish, thereby increasing the accuracy and efficiency of the expansion. For $\zeta$-quadratic Hamiltonians the classical approximation is exact, namely $\Gamma_{s,t} = \gamma_{s,t}^{(0)}$.

Prior literature on the semiclassical expansion (1.8) of the Heisenberg–Weyl evolution operator is evidently sparse. However, for the expansion of the Weyl symbol of a time evolving observable, as in equation (3.31), there are a variety of results. The conceptual ingredients (found in Section 4) of employing transport recurrence relations obtained from an order-by-order $\hbar$ expansion of the Moyal equation of motion goes back to Berezin and Shubin.[9,11] When applied to densities rather than observables, the recurrence transport method has been treated extensively by Prosser.[10,39]

The most widely studied semiclassical expansion in the literature is the WKB approximation for the propagator $\langle x | U(t,s) | y \rangle$. As such, it establishes a natural benchmark for measuring the effectiveness of the Heisenberg–Weyl expansion (1.8). In the expansion (1.8) there is only one classical trajectory determined by an initial value problem. The coefficients $\gamma_{s,t}^{(n)}$ are singularity free and, if the classical motion does not run away to infinity in finite time, are defined for all time. Unlike the WKB approximation there is no essential singularity in $\hbar$, no need to solve a two-point variational problem for the relevant classical trajectories, and no small time restrictions or caustic difficulties.

It was emphasized in §3 that this globally unique trajectory aspect of expansion (1.8) is the joint result of employing Weyl symbols for Heisenberg picture evolution. It is worthwhile to have a somewhat heuristic understanding of this theme. Certainly the use of phase space based symbols (such as Weyl's) admits the possibility of a unique trajectory being the relevant one (in the Weyl case its initial value is $\zeta$, the symbol variable). Such a possibility seems naturally excluded, e.g., for the usual $\langle x | \cdots | y \rangle$ representation because the pair of configuration variables $(x,y)$ is not equivalent to a point in phase space. However, use of the Weyl symbol alone in studying quantum evolution is not sufficient to guarantee absence of caustics. For example, if the Schrödinger picture is used and $U_{\mathrm{w}}(t,s)$ is studied, the relevant boundary condition[38,40] for trajectories is $\big(z_c(t) + z_c(s)\big)/2 = \zeta$, which admits caustics and carries no global guarantee of unique solutions. Why is the situation so favorably different for the Heisenberg picture evolution operator $\boldsymbol{\Gamma}(s,t)$? Roughly, this seems to be because, through the appearance of both $U^\dagger$ and $U$, the Heisenberg operator $\boldsymbol{\Gamma}(s,t)A$ maps between states at a common time, $s$, which turns out to be the initial time for the trajectory value $\zeta$.

## APPENDIX A: AFFINE CANONICAL COVARIANCE

The Wigner–Weyl picture of quantum mechanics possesses a simple covariance with respect to affine canonical transformations.[11,16,20,26,73,74] The first part of this appendix reviews the nature of this covariance, while the second part examines precisely how this property is respected by the formulae and methods of computation derived in §3.

Let $\Phi : \mathbb{R}^{2d} \to \mathbb{R}^{2d}$ be an affine canonical transformation. For convenience we represent $\Phi$ as

$$\Phi(z) = \mathcal{S}^{-1}(z - z_o), \qquad (A.1a)$$

or equivalently

$$\Phi^{-1}(\zeta) = \mathcal{S}\zeta + z_o, \qquad (A.1b)$$

where the symplectic matrix $\mathcal{S} \in Sp(2d)$ and the displacement $z_o \in \mathbb{R}^{2d}$ are arbitrary. Let us say that a unitary operator $V$ *corresponds to* $\Phi$ if the following quantum counterpart of $(A.1a)$ holds,

$$V \hat{z} V^\dagger = \Phi(\hat{z}) \equiv \mathcal{S}^{-1}(\hat{z} - z_o \hat{I}). \qquad (A.2a)$$

In this case

$$V^\dagger \hat{z} V = \mathcal{S}\hat{z} + z_o \hat{I}. \qquad (A.2b)$$

Given $\Phi$, we note that such corresponding operators $V = V(\mathcal{S}, z_o)$ always exist. Indeed they may be represented by $V = T(z_o) M(\mathcal{S})$ where $T(z_o)$ is the Heisenberg phase space translation operator[16,75,76]

$$T(z_o) \equiv \exp\left(\frac{i}{\hbar} z_o \cdot J^{-1} \hat{z}\right), \qquad (A.3)$$

and $M(\mathcal{S})$ is a metaplectic operator.[16,26,77] These operators respectively obey the fundamental identities $T(z_o)^\dagger \hat{z} T(z_o) = \hat{z} + z_o \hat{I}$ and $M(\mathcal{S})^\dagger \hat{z} M(\mathcal{S}) = \mathcal{S}\hat{z}$. Condition $(A.2a)$ may be shown to determine $V$ uniquely up to a phase (i.e. if $V$, $V'$ obey $(A.2a)$ then $V = cV'$ with $|c| = 1$).

In terms of such corresponding classical and quantum affine transformations, the covariance of the Wigner transform takes the form:

$$\left(V^\dagger \hat{A} V\right)_w = A_w \circ \Phi^{-1} \qquad (A.4)$$

for any operator $\hat{A}$. Before proving this identity, it is of interest to clarify its interpretation.

Regard $\Phi$ as a (passive) change of phase space coordinates, $z \mapsto \zeta \equiv \Phi(z)$. If $A_w$ is a classical observable, then $A_w \circ \Phi^{-1}$ represents the same observable in the new coordinates. On the other hand, the unitary operator $V$ can be used to implement a change of basis vectors in $\mathcal{H}$, defined by $\psi \mapsto \phi \equiv V^\dagger \psi$. Under this 'quantum coordinate change' the observable $\hat{A}$ transforms into $V^\dagger \hat{A} V$. Thus $(A.4)$ says that, for corresponding transformations, the coordinate change process commutes with the Wigner–Weyl mapping between quantum and classical observables.

We now give the proof of $(A.4)$. Conjugating the representation $(2.4)$ of $\hat{A}$ in terms of its symbol, with $V$, results in

$$V^\dagger \hat{A} V = (2\pi)^{-2d} \int du \int d\zeta \, A_w(\zeta) \, e^{iu \cdot (\mathcal{S}\hat{z} + z_o - \zeta)}.$$

Upon making the change of integration variables $v = \mathcal{S}^\mathrm{T} u$ and $w = \mathcal{S}^{-1}(\zeta - z_o)$ one finds

$$V^\dagger \hat{A} V = (2\pi)^{-2d} \int dv \int dw\, A_\mathrm{w}(\mathcal{S}w + z_o)\, e^{iv\cdot(\hat{z}-w)},$$

which implies $(A.4)$.

Let us now consider the covariance of the semiclassical method developed in §3. For brevity, denote conjugation by a unitary $V$ with a superscript, $\hat{A}^V \equiv V^\dagger \hat{A} V$. From a general perspective, the transformation properties of all semiclassical symbols are immediately given by $(A.4)$; thus

$$\left(\hat{A}(t,s)^V\right)_\mathrm{w}(\zeta) = A_\mathrm{w}\left(t,s,\hbar;\Phi^{-1}(\zeta)\right). \qquad (A.5)$$

Taking the example $\hat{A} = \hat{z}$ and combining this with the $\hbar$ expansion (3.3) yields

$$\left(\hat{Z}(t,s)^V\right)_\mathrm{w}(\zeta) = \sum_{r=0}^{\infty} \frac{\hbar^r}{r!}\, \mathbf{z}_r\left(t,s;\Phi^{-1}(\zeta)\right). \qquad (A.6)$$

Assume that the affine canonical map $\Phi$, corresponding to $V$, is $\hbar$ free (i.e. that $\mathcal{S}$ and $z_o$ don't depend on $\hbar$), as is natural from a semiclassical perspective. Then $(A.6)$ will be a *bona fide* $\hbar$ expansion of the form

$$Z_\mathrm{w}^V(t,s,\hbar;\zeta) = \sum_{r=0}^{\infty} \frac{\hbar^r}{r!}\, \mathbf{z}_r^V(t,s;\zeta), \qquad (A.7a)$$

with

$$\mathbf{z}_r^V(t,s;\zeta) = \mathbf{z}_r\left(t,s;\Phi^{-1}(\zeta)\right) \qquad (A.7b)$$

being the covariance relation for the coefficient functions. In the remainder of this appendix, we indicate how property $(A.5)$ emerges if $\hat{A}(t,s)^V{}_\mathrm{w}$ is computed by the methods of §3, assuming that $(A.7)$ is given. In this regard, we remark that it is possible to derive $(A.7b)$, independently of the above discussion, by examining the Jacobi equations obeyed by $\mathbf{z}_r^V$ in an inductive fashion. We will not present this inductive proof since it is tedious and several of its facets would be duplicated by the proof of $(A.5)$ which we sketch next.

To begin the derivation, conjugate the operator equation (3.32) with $V$,

$$\hat{A}(t,s)^V = \exp\left\{\hat{Z}(t,s)^V \cdot \nabla\right\} A_\mathrm{w}(t,\hbar;0). \qquad (A.8)$$

Note that it is $A_\mathrm{w}$ [not $(\hat{A}^V)_\mathrm{w}$] which appears on the right side, because operators [not symbols] are being conjugated. Using the cluster theorem, the Wigner transform of $(A.8)$ is

$$A_\mathrm{w}^V(t,s,\hbar;\zeta) = e^{\delta Z_\mathrm{w}^V(t,s,\hbar;\zeta)\cdot\nabla} \exp\left\{\sum_{j=2}^{\infty} \frac{1}{j!}\mathcal{L}_j^V(\zeta;\hbar,t,s,\nabla)\right\} A_\mathrm{w}(t,\hbar;\mathbf{z}_0^V(t,s|\zeta)), \qquad (A.9)$$

in which the relevant cluster operators are, cf. (3.8b),

$$\mathcal{L}_j^V(\zeta;\hbar,t,s,\nabla) = \sum_{C\in\mathcal{C}_j} \prod_{\alpha\in E(C)} \left[e^{(i\hbar/2)B_\alpha} - 1\right] \underset{j}{\times} Z_{\mathrm{w}}^V(t,s,\hbar;\zeta)\cdot\nabla. \quad (A.10)$$

Consider a typical extended bracket $B_\alpha = \nabla_k \cdot J\nabla_l$ acting on the $k^{\mathrm{th}}$ and $l^{\mathrm{th}}$ instances of $Z_{\mathrm{w}}^V$ in the product function. Notice that (A.7) implies at once the version of (A.5) restricted to the case $\hat{A} = \hat{z}$, viz.

$$Z_{\mathrm{w}}^V(t,s,\hbar;\zeta) = Z_{\mathrm{w}}\bigl(t,s,\hbar;\Phi^{-1}(\zeta)\bigr).$$

Using this identity to replace each $Z_{\mathrm{w}}^V$ in (A.10) leads, by the chain rule, to the replacement of $B_\alpha$ with

$$\nabla_k \cdot \nabla\Phi^{-1}(\zeta) J \nabla\Phi^{-1}(\zeta)^{\mathrm{T}} \nabla_l, \quad (A.11)$$

where now $\nabla_k$ acts on the original symbol $Z_{\mathrm{w}}(t,s,\hbar;\,\cdot\,)$, followed by evaluation at $\Phi^{-1}(\zeta)$. Since $\Phi^{-1}$ is canonical ($\nabla\Phi^{-1} = \mathcal{S}$ is symplectic) the bracket (A.11) reduces to precisely $B_\alpha$. Moreover, since $\Phi^{-1}$ is affine, the derivative $\mathcal{S}$ is $\zeta$ independent and so the above bracket invariance extends to arbitrary higher order derivatives arising from any $e^{(i\hbar/2)B_\alpha}$ in (A.10). This results in the covariance rule for the cluster operators,

$$\mathcal{L}_j^V(\zeta;\hbar,t,s,\nabla) = \mathcal{L}_j\bigl(\Phi^{-1}(\zeta);\hbar,t,s,\nabla\bigr). \quad (A.12)$$

Finally, use of (A.7b) and (A.12) to replace the $V$ dependent objects on the right side of (A.9) completes the verification of (A.5). The $\hbar$ coefficient covariance $\mathbf{a}_r^V = \mathbf{a}_r \circ \Phi^{-1}$ follows from this.

### APPENDIX B: INTEGRAL KERNEL REPRESENTATIONS

Like most quantum operators, the Heisenberg–Weyl evolution $\boldsymbol{\Gamma}(s,t)$ has an integral kernel realization. In this appendix we present two established phase space representations of Heisenberg–Weyl evolution and clarify their linkage to $\Gamma_{s,t}$.

The kernel representation of $\Gamma_{s,t}$ is

$$\Gamma_{s,t}(a)(z) = \int_{\mathbb{R}^{2d}} d\zeta\, K(s,z|t,\zeta) a(\zeta). \quad (B.1)$$

Formally, $K$ is the distribution valued solution of (1.12) which has the $s = t$ delta function initial value $\delta_z(\zeta)$. The ordering of the time labels in (B.1) reflects the fact that $s$ is the active time variable in the backward equation of motion (4.4a) and that $t$ is the initial pinning time. Given the existence of $K$, setting $a(\zeta) = \delta_{z'}(\zeta)$ shows that

$$\Gamma_{s,t}(\delta_{z'})(z) = K(s,z|t,z'). \tag{B.2}$$

Formula $(B.2)$ provides the link between the $\Gamma_{s,t}$ operator and the kernel $K$. Putting $a = \delta_z$ into (3.37)–(3.38) explicitly displays the distributional nature of the formal $\hbar \downarrow 0$ asymptotic expansion of $K$.

One standard form of $K$ is realized in terms of $\Delta(z)$, the quantized delta function operator[78,79] defined by $\Delta(z)_\mathrm{w} = h^d \delta_z$. The work of Grossmann[80] and Royer[54] has shown that $\Delta$ is essentially the product of two elementary unitary operators, namely $\Delta(z) = 2^d T(2z)\Pi$. Here $T$ is the Heisenberg translation operator $(A.3)$, and $\Pi$ is the parity operator, $\Pi \hat{z} = -\hat{z}\Pi$. The usefulness of $\Delta$ is apparent from the identities

$$\hat{A} = \int \frac{d\zeta}{h^d} a(\zeta)\Delta(\zeta), \qquad a(z) = \mathrm{Tr}\bigl[\hat{A}\Delta(z)\bigr]. \tag{B.3a,b}$$

The first of this pair provides another way to Weyl quantize the symbol $a$, comparable to (2.4) or (2.5). The second identity determines the Weyl symbol $a$ from the operator $\hat{A}$ via the trace, and is an alternative to the Wigner transform (2.7). Straightforward demonstrations of $(B.3)$ may be found in Ref. [53].

We now review an exact representation[36,53] of $K(s,z|t,\zeta)$. Employing $(B.3a)$ allows one to write the Heisenberg evolution of $\hat{A}$ as

$$\boldsymbol{\Gamma}(s,t)\hat{A} = \int \frac{d\zeta}{h^d} U(t,s)^\dagger \Delta(\zeta) U(t,s) A_\mathrm{w}(\zeta). \tag{B.4}$$

The symbol of $\boldsymbol{\Gamma}(s,t)\hat{A}$ is $\Gamma_{s,t}(A_\mathrm{w})(z)$. A second form of this symbol is obtained from the right side of $(B.4)$ by multiplying by $\Delta(z)$ and taking the trace to give

$$K(s,z|t,\zeta) = h^{-d}\mathrm{Tr}\bigl[U(t,s)^\dagger \Delta(\zeta) U(t,s)\Delta(z)\bigr]. \tag{B.5}$$

This formula states that kernel $K(s,z|t,\zeta)$ is the expectation value that results from the $s \to t$ evolution of observable $\Delta(\zeta)/h^d$ with respect to the 'density matrix' $\Delta(z)$. The kernel $K$ is real valued, and satisfies the symmetry condition $K(s,z|t,\zeta) = K(t,\zeta|s,z)$, as is easy to show from $(B.5)$. It also obeys a composition law implied by (4.5).

As the recent work of Marinov[81] has revealed, $K$ admits a path integral representation. For a Weyl quantized Hamiltonian, the standard time slicing method in combination with the short time approximation $U_\mathrm{w}(\tau + \Delta t, \tau)(z) \approx \exp\bigl(-i\Delta t H_\mathrm{c}(\tau,z)/\hbar\bigr)$ leads to

$$K(s,z|t,\zeta) = \frac{1}{h^d} \int D\tilde{z}D\tilde{\xi}\, e^{(i/\hbar)\mathcal{S}(\tilde{z},\tilde{\xi})}.$$

Here $D\tilde{z}$ and $D\tilde{\xi}$ are infinite products of the dimensionless measures $h^{-d}dz$ and $h^{-d}d\xi$. The phase is the action-like functional

$$\mathcal{S}(\tilde{z},\tilde{\xi}) = \int_s^t d\tau \left[ \tilde{\xi}(\tau) \cdot J\dot{\tilde{z}}(\tau) + H_c\big(\tau, \tilde{z}(\tau) + \tfrac{1}{2}\tilde{\xi}(\tau)\big) - H_c\big(\tau, \tilde{z}(\tau) - \tfrac{1}{2}\tilde{\xi}(\tau)\big) \right].$$

The phase space paths $\tilde{z}(\,\cdot\,)$ obey the boundary conditions $\tilde{z}(s) = z$, $\tilde{z}(t) = \zeta$, whereas the variations $\tilde{\xi}(\,\cdot\,)$ are unrestricted.

## Acknowledgements

This research was supported in part by grants to **TAO** from the Natural Sciences and Engineering Research Council of Canada. The authors are grateful to Steve Fulling (Texas A&M) for critically reading the manuscript and suggesting a number of improvements, and thank Andrei Barvinsky (Moscow Nuclear Safety Institute) for a helpful discussion.

## References

1. J. E. Moyal, *Proc. Cambr. Phil. Soc.* **45** (1949), 99.
2. E. P. Wigner, *Phys. Rev.* **40** (1932), 749.
3. J. C. T. Pool, *J. Math. Phys.* **7** (1966), 66.
4. K. İmre, E. Özizmir, M. Rosenbaum and P. F. Zweifel, *J. Math. Phys.* **8** (1967), 1097.
5. E. P. Wigner, *in* "Perspectives in Quantum Theory" (W. Yourgrau and A. van der Merwe, Eds.), p. 25, MIT Press, Cambridge, MA, 1971.
6. R. F. O'Connell and E. P. Wigner, *Phys. Lett.* **83A** (1981), 145.
7. M. Hillery, R. F. O'Connell, M. O. Scully and E. P. Wigner, *Phys. Rep.* **106** (1984), 121.
8. F. J. Narcowich, *in* "Wigner Distribution Functions," Seminars in Mathematical Physics, No. 1, p. 1, Texas A&M Univ., College Station, TX, 1986.
9. F. A. Berezin and M. A. Šubin, *in* "Proceedings of the Colloquia Mathematica Societatis János Bolyai, 5. Hilbert Space Operators," p. 21, North-Holland, Amsterdam, 1972.
10. R. T. Prosser, *J. Math. Phys.* **24** (1983), 548.
11. F. A. Berezin and M. A. Shubin, "The Schrödinger Equation," Klewer Academic, Dordrecht, 1991.
12. H. J. Groenewold, *Physica* **12** (1946), 405.

13. R. Abraham, J. E. Marsden and T. Ratiu, "Manifolds, Tensor Analysis, and Applications," Addison–Wesley, Reading, MA, 1983.
14. M. V. Karasev and V. E. Nazaikinskii, *Math. USSR-Sb.* **34** (1978), 737.
15. N. L. Balazs and B. K. Jennings, *Phys. Rep.* **104** (1984), 347.
16. R. G. Littlejohn, *Phys. Rep.* **138** (1986), 193.
17. H. Weyl, *Zeitschr. f. Phys.* **46** (1927), 1.
18. N. H. McCoy, *U. S. Natl. Acad. Sci.* **18** (1932), 674.
19. A. Grossmann, G. Loupias and E. M. Stein, *Ann. Inst. Fourier* **18** (1968), 343.
20. L. Hörmander, *Comm. Pure Appl. Math.* **32** (1979), 359.
21. J. M. Gracia-Bondía and J. C. Várilly, *J. Math. Phys.* **29** (1988), 869.
22. J. C. Várilly and J. M. Gracia-Bondía, *J. Math. Phys.* **29** (1988), 880.
23. M. A. Antonets, *Moscow Univ. Math. Bull.* **32(6)** (1977), 21.
24. A. Voros, *J. Func. Anal.* **29** (1978), 104.
25. R. Estrada, J. M. Gracia-Bondía and J. C. Varilly, *J. Math. Phys.* **30** (1989), 2789.
26. G. B. Folland, "Harmonic Analysis in Phase Space," Princeton Univ. Press, Princeton, NJ, 1989.
27. M. A. Shubin, "Pseudodifferential Operators and Spectral Theory," Springer-Verlag, Berlin, 1986.
28. F. Bayen, M. Flato, C. Fronsdal, A. Lichnerowicz and D. Sternheimer, *Ann. Phys. (N.Y.)* **111** (1978), 61.
29. F. Bayen, M. Flato, C. Fronsdal, A. Lichnerowicz and D. Sternheimer, *Ann. Phys. (N.Y.)* **111** (1978), 111.
30. R. M. Wilcox, *J. Math. Phys.* **8** (1967), 962.
31. D. B. Fairlie and C. M. Manogue, *J. Phys. A: Math. Gen.* **24** (1991), 3807.
32. I. S. Gradshteyn and I. M. Ryzhik, "Table of Integrals, Series, and Products," Academic, San Diego, 1965.
33. R. J. Wilson, "Introduction to Graph Theory," Chap. 2, Academic, New York, 1975.
34. T. L. Hill, "Statistical Mechanics," McGraw–Hill, New York, 1956.
35. F. H. Molzahn and T. A. Osborn, *Ann. Phys. (NY)* **230** (1994), 343.
36. T. B. Smith, *J. Phys. A* **11** (1978), 2179.
37. A. Intissar, *Comm. Par. Diff. Eq.* **7** (1982), 1403.
38. M. V. Berry, *Proc. Roy. Soc. Lond.* **A 423** (1989), 219.
39. R. T. Prosser, *On the Correspondence between Classical and Quantum Mechanics. II*, Dartmouth College preprint, 1993.
40. M. S. Marinov, *J. Phys. A* **12** (1979), 31.
41. T. A. Osborn and F. H. Molzahn, *in* "Forty More Years of Ramifications: Spectral Asymptotics and its Applications" (S. A. Fulling and F. J. Narcowich, Eds.),

Discourses in Mathematics and its Applications No. 1, p. 199, Texas A&M Univ. Math. Dept., College Station, TX, 1991.
42. M. E. TAYLOR, "Pseudodifferential Operators," Princeton Univ. Press, Princeton, NJ, 1981.
43. P. CARRUTHERS AND F. ZACHARIASEN, *Rev. Mod. Phys.* **55** (1983), 245.
44. R. BALIAN AND M. VÉNÉRONI, *Ann. Phys. (NY)* **164** (1985), 334.
45. R. BALIAN AND M. VÉNÉRONI, *Nucl. Phys.* **B408** (1993), 445.
46. F. J. NARCOWICH, *J. Math. Phys.* **27** (1986), 2502.
47. S. G. KREIN, "Linear Differential Equations in Banach Space," Am. Math. Soc., Providence, RI, 1971.
48. A. SAKSENA, T. A. OSBORN AND F. H. MOLZAHN, *J. Math. Phys.* **32** (1991), 938.
49. T. KATO, "Perturbation Theory for Linear Operators," Springer-Verlag, Berlin, 1966.
50. R. SCHATTEN, "Norm Ideals of Completely Continuous Operators," Springer-Verlag, Berlin, 1960.
51. G. A. BAKER, JR., *Phys. Rev.* **109** (1958), 2198.
52. G. LOUPIAS AND S. MIRACLE-SOLE, *Ann. Inst. Henri Poincaré* **6** (1967), 39.
53. S. R. DEGROOT AND L. G. SUTTORP, "Foundations of Electrodynamics," North-Holland, Amsterdam, 1972.
54. A. ROYER, *Phys. Rev.* **A 15** (1977), 449.
55. J. G. KIRKWOOD, *Phys. Rev.* **44** (1933), 31.
56. Y. FUJIWARA, T. A. OSBORN AND S. F. J. WILK, *Phys. Rev.* **A 25** (1982), 14.
57. T. A. OSBORN, *J. Phys.* **A 17** (1984), 3477.
58. L. PAPIEZ, T. A. OSBORN AND F. H. MOLZAHN, *J. Math. Phys.* **29** (1988), 642.
59. F. J. NARCOWICH, *Physica* **134A** (1985), 193.
60. M. V. FEDORIUK, *Russian Math. Surv.* **26** (1971), 65.
61. E. J. HELLER, *J. Chem. Phys.* **65** (1976), 1289.
62. M. V. BERRY, *Phil. Trans. Roy. Soc. Lond.* **A 287** (1977), 237.
63. C. MORETTE, *Phys. Rev.* **81** (1951), 848.
64. V. P. MASLOV, *USSR Comput. Math. Math. Phys.* **1** (1962), 123.
65. D. W. MCLAUGHLIN, *J. Math. Phys.* **13** (1972), 784.
66. M. V. BERRY AND K. E. MOUNT, *Rep. Prog. Phys.* **35** (1972), 315.
67. M. M. MIZRAHI, *J. Math. Phys.* **19** (1978), 298.
68. M. M. MIZRAHI, *J. Math. Phys.* **22** (1981), 102.
69. B. SIMON, "Functional Integration and Quantum Physics," Academic, New York, 1979.
70. V. P. MASLOV AND M. V. FEDORIUK, "Semiclassical Approximation in Quantum Mechanics," Reidel, Dordrecht, 1981.
71. M. C. GUTZWILLER, *J. Math. Phys.* **8** (1967), 1979.
72. S. L. ROBINSON, *J. Math. Phys.* **29** (1988), 412.

73. N. L. Balazs, *Physica* **102A** (1980), 236.
74. F. J. Narcowich, *J. Math. Phys.* **31** (1990), 354.
75. J. R. Klauder and E. C. G. Sudarshan, "Fundamentals of Quantum Optics," Benjamin, New York, 1968.
76. M. E. Taylor, "Noncommutative Harmonic Analysis," Math. Surv. Mon. 22, American Mathematical Society, Providence, RI, 1986.
77. M. Moshinsky and C. Quesne, *J. Math. Phys.* **12** (1971), 1772.
78. B. Leaf, *J. Math. Phys.* **9** (1968), 65.
79. J. F. Cariñena, J. M. Gracia-Bondía and J. C. Várilly, *J. Phys. A: Math. Gen.* **23** (1990), 901.
80. A. Grossmann, *Commun. Math. Phys.* **48** (1976), 191.
81. M. S. Marinov, *Phys. Lett. A* **153** (1991), 5.